# Transverse optical gradient force in untethered rotating metaspinners

Einstom Engay[1,#], Mahdi Shanei[1,#], Vasilii Mylnikov[1], Gan Wang[2], Peter Johansson[3], Giovanni Volpe[2], and Mikael Käll[1*]

[1]Department of Physics, Chalmers University of Technology, 412 96 Gothenburg Sweden

[2]Department of Physics, University of Gothenburg, 412 96 Gothenburg, Sweden

[2]School of Science and Technology, Örebro University, 701 82 Örebro, Sweden

#Equal contributions.

*Corresponding author: mikael.kall@chalmers.se

**Abstract.** Nanostructured dielectric metasurfaces offer unprecedented opportunities to control light-matter momentum exchange, and thereby the forces and torques that light can exert on matter. Here we introduce optical metasurfaces as components of ultracompact untethered microscopic *metaspinners* capable of efficient light-induced rotation in a liquid environment. Illuminated by weakly focused light, a metaspinner generates torque via photon recoil through the metasurfaces' ability to bend light towards high angles despite their sub-wavelength thickness, thereby creating orbital angular momentum. We find that a metaspinner is subject to an anomalous transverse lateral optical gradient force that acts in concert with the classical gradient force. Consequently, when two or more metaspinners are trapped together in a laser beam, they collectively orbit the optical axis in the opposite direction to their spinning motion, in stark contrast to rotors coupled through hydrodynamic or mechanical interactions. The metaspinners delineated herein not only serve to illustrate the vast possibilities of utilizing optical metasurfaces for fundamental exploration of optical torques, but they also represent potential building-blocks of artificial active matter systems, light-driven micromachinery, and general-purpose optomechanical devices.

## Introduction

Micro and nanostructures amenable to non-contact manipulation using external driving fields receive increasing attention across the natural sciences as a consequence of the possibility to generate novel fundamental understandings of mesoscopic phenomena[1,2] and the possibility for novel applications within areas such as biomedicine[3-5], material delivery[6], and micro/nanomechanical systems[7-9]. For example, remotely actuated microstructures have been used to emulate the behavior of biological microorganisms[10,11] while synthetic microstructures operated as untethered micromechanical tools have been explored as micropumps[12], microgrippers[13,14], and hydrodynamic propulsion systems[15,16].

Among the various field-actuation mechanisms available, including magnetic[1,13], electric[8,17,18], and acoustic[19], microparticle manipulation based on optical forces is particularly appealing due to the versatility, maturity and widespread availability of sophisticated optical microscopes and laser systems. Optical tweezers, which are based on conservation of linear momentum in the light-matter interaction, remain the prime example in this category [20,21]. However, there is also the possibility to optically rotate particles, in this case based on conservation of angular momentum[22,23]. This phenomenon was long seen as a mere scientific curiosity but developments since the mid 1990's[24] have demonstrated that optical rotation is a powerful tool for fundamental studies as well as applications[25-28]. Consequently, significant research efforts have been devoted to development of micro- and nanostructures amenable to efficient optical rotation[29-33].

A light beam exerts torque on an object if the object changes the spin angular momentum (SAM) and/or the orbital angular momentum (OAM) of the beam, as determined by its polarization properties and azimuthal distribution of linear momentum, respectively[34-36]. Examples of objects designed to be efficiently rotated through SAM transfer include vaterite microspheres[29] and plasmonic nanorods[32] whereas efficient OAM transfer is more difficult to achieve since the object needs to be engineered to deflect light in specific directions[37-39]. A sophisticated example of the latter is the "optical reaction micro-turbine", designed to redirect an incident focused laser beam via a bundle of twisted light guides[40].

Recently, ultrathin microparticles supporting nanostructures designed to generate optical forces have emerged as a promising concept for realizing versatile optical manipulation schemes[41-43]. High-index dielectric metasurfaces are particularly attractive in this context because of the possibility to modulate the amplitude, phase, and polarization of transmitted and reflected light with unprecedented control and flexibility in an ultrathin and low-drag format[42,44,45]. Furthermore, the possibility to produce thousands of individual metastructures in a single lithographic process, and the possibility to activate these using essentially unfocused and unpolarized light, open new avenues of applications in optically driven micromachinery and for the development of various kinds of microrobotic and microfluidic applications.

Here we demonstrate a novel approach in constructing untethered and freely moving microrotors based on optical metasurfaces embedded in transparent host particles. Each "metaspinner" contains a pair of metagratings that deflect light in opposite directions via a "lever arm", thereby creating OAM that allows the particles to be efficiently rotated under loosely focused illumination. We first analyze the dynamics of individual metaspinners; then we study the behavior of multiple interacting particles, which reveals a counterintuitive collective orbital motion in the opposite direction to the spinning motion. This anomalous behavior originates in a transverse lateral gradient force that dominates hydrodynamic and mechanical interactions. We finally discuss possible further developments and potential applications of the results described herein.

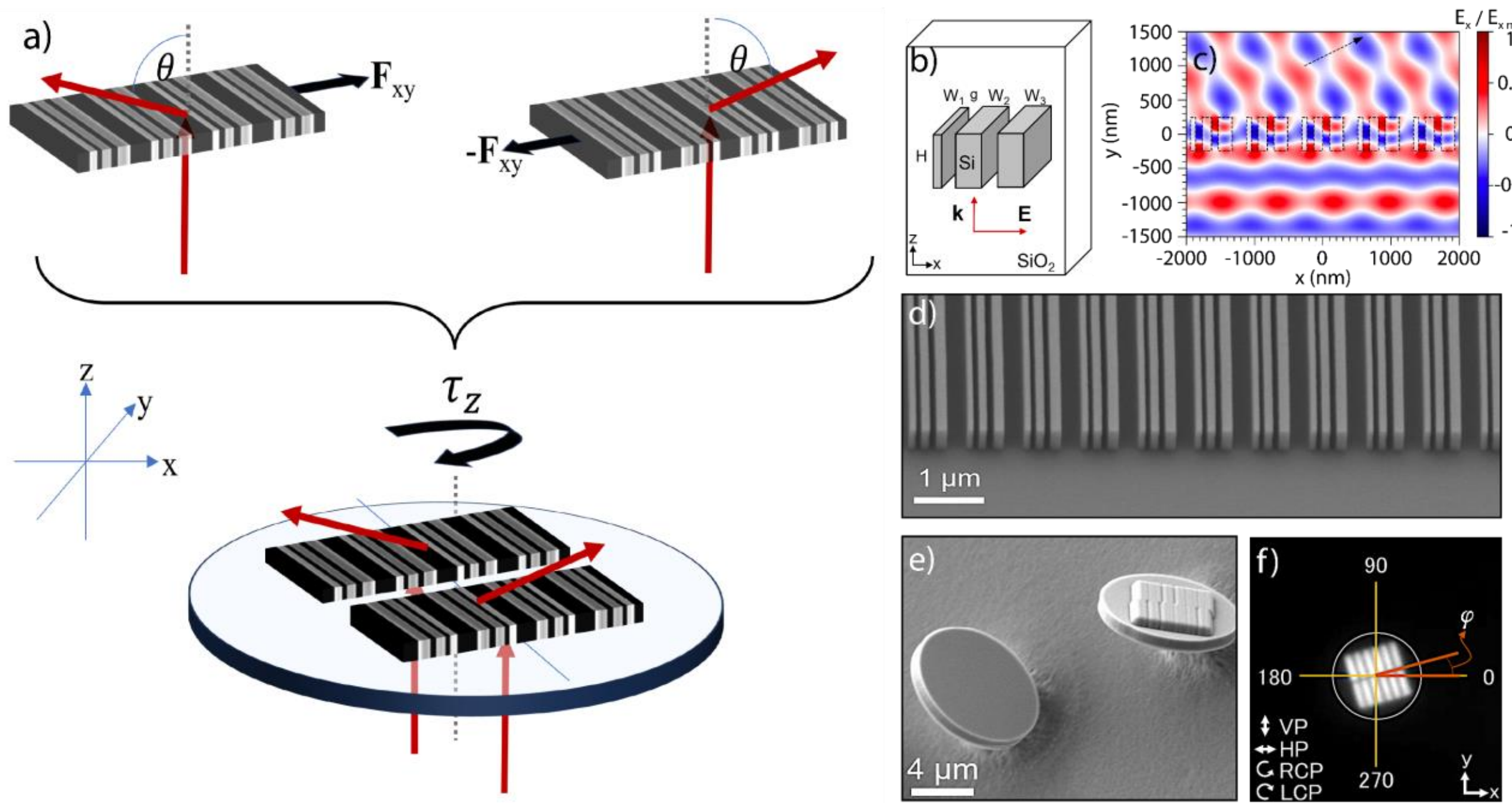

**Figure 1. Metaspinner concept and design.** (a) A metaspinner contains two identical metagratings oriented to deflect transmitted light in opposite directions at a high angle $\theta$, thereby inducing an optical torque $\tau_z = 2\mathbf{F}_{xy} \times \mathbf{r}$ that forces the spinner to rotate. (b) The optimized metagrating unit cell contains three amorphous silicon ridges with varying widths ($W_1$ = 90 nm, $W_2$ = 130 nm, $W_3$ = 190 nm) but constant height ($H$ = 490 nm), gap width ($g$ = 100 nm), and periodicity ($\Lambda$ = 817 nm). (c) Simulated electric-field distribution around an optimized metagrating illuminated with $\lambda$ = 1064 nm plane-waves at normal incidence, indicating the efficient bending of p-polarized

light towards $\theta \approx 64^o$, which corresponds to the transmitted +1-diffraction order. (d) SEM image of a fabricated metagrating and of (e) metaspinners ready for release; (f) Reflected bright field optical image of a metaspinner indicating the azimuthal angle $\varphi$ between the diffraction plane and the laboratory xz-plane. The incident polarization states indicated with arrows correspond to vertical polarization (VP), horizontal polarization (HP), right-handed circular (RCP) and left-handed circular polarization (LCP). The white circle indicates the border of the metaspinner.

## Results

### Metaspinner concept and design.

The basic idea behind the metaspinner is schematically illustrated in Fig. 1a: A single metasurface able to deflect light incident normal to its surface experiences a reaction force due to conservation of linear momentum. The in-plane component of this photon-recoil force, $\mathbf{F}_{xy}$, is determined by the deflection angle $\theta$, the deflection efficiency, and by the incident power, and can be used to propel the metasurface within the transverse plane[40]. If two such metasurfaces are oriented in opposite directions and then joined together in a single rigid body, the reaction forces instead generate a torque $\tau_z = 2\mathbf{F}_{xy} \times \mathbf{r}$ because of the non-zero lever arms $\mathbf{r}$ to the common center of mass (c.o.m). This torque is due to optical orbital angular momentum transfer, since it does not involve the intrinsic angular momentum (spin) of light, and it can be used to rotate the metaspinner around its c.o.m.

We fabricated 8 µm diameter metaspinners with a thickness of ~1 µm by a combination of electron beam lithography and etching (see Methods). Each metaspinner contains two oppositely oriented 4.65 x 2.5 µm$^2$ amorphous silicon (aSi) metagratings embedded in $SiO_2$. The unit cell (Fig. 1b) was optimized for p-polarized directional diffraction at a wavelength of $\lambda$ = 1064 nm using electrodynamic simulations and measurements on large samples (see Methods and Supporting Fig. S1-S2). An optimized metagrating contains three aSi ridges per unit cell and has a periodicity corresponding to a deflection angle of $\theta = 64^o$ in $SiO_2$. Figure 1c shows a calculated near-field profile for this case, illustrating the effective bending of an incident plane wave towards the main diffraction direction (defined as the +1 transmitted order) while Fig. 1e shows a scanning electron microscope (SEM) image of a fabricated sample. The structures are two-dimensionally chiral (point group $C_{2h}$), which implies that a metaspinner can be designed to rotate either clockwise (CW, left-handed (LH) structure) or counterclockwise (CCW, right-handed (RH) structure), but the handedness reverses if a metaspinner is turned upside down. Although each metagrating in a spinner only contains six periods of the unit cell, we find that the resulting orbital torque suffices to cause a strong spinning motion.

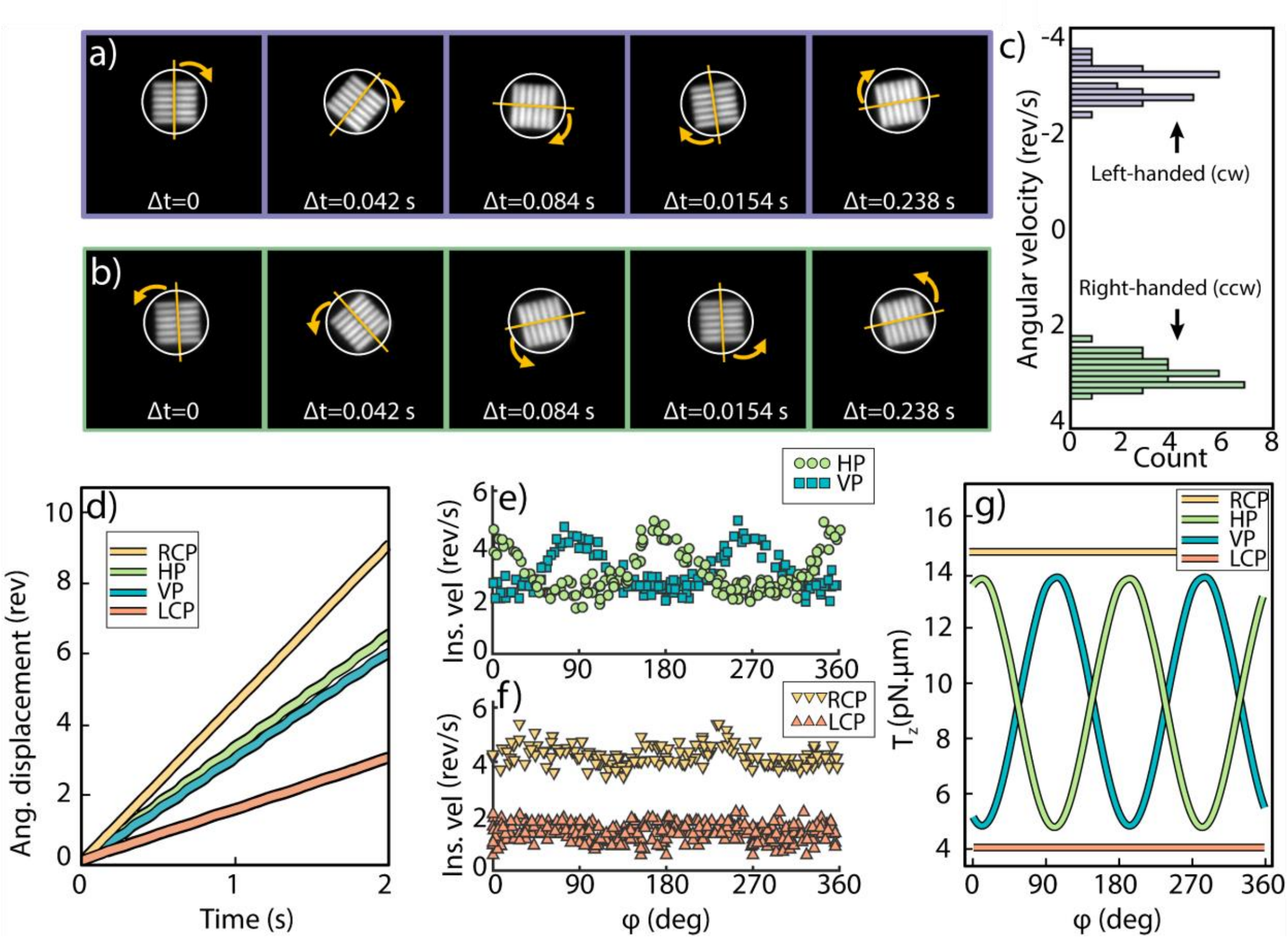

**Figure 2. Rotation dynamics of individual metaspinners.** (a) Left-handed (LH) and (b) right-handed (RH) metaspinners exhibiting clockwise (CW) and counterclockwise (CCW) rotation, respectively, when exposed to linearly polarized light. The periphery of the $SiO_2$ disks supporting the metagratings are outlined by white circles. See also Supporting Videos 1 and 2. (c) Histogram of rotation frequencies for several LH and RH metaspinners. (d) Angular displacement $\varphi(t)$ of the RH metaspinner for RCP, LCP, HP, and VP incident fields. Here $\varphi$ signifies the angle between the metagrating diffraction plane and the laboratory x-axis, as indicated in Fig. 1f). (e, f) Angular speed versus orientation $\dot{\varphi}(\varphi)$ for the four polarization cases, obtained from the data in (d). (g) Calculated total optical torque versus orientation $\tau_{opt}(\varphi)$ for the four polarization cases from FDTD simulations and Maxwell stress tensor analysis of a RH metaspinner. The experimental data in (a-f) were obtained using a loosely focused $\lambda = 1064$ nm laser beam with peak intensity ~75 μW/μm$^2$. The calculated torque in (g) was obtained for plane wave incidence and an intensity of 75 μW/μm$^2$.

## Rotation dynamics of individual metaspinners.

After being released from the substrate, the metaspinners were dispersed in water and a droplet of the solution was injected into a thin liquid cell mounted on an inverted microscope (Supporting Fig. S3). The driving laser beam was loosely focused to a Gaussian spot (FWHM = 25.4 μm) and controlled in power and polarization. For sufficient power, the classical optical gradient force was found strong enough to enable 2D trapping and lateral manipulation of metaspinners that had sedimented at the bottom of the liquid cell. Figure 2a and 2b (Supporting Videos 1 and 2) show examples of LH and RH metaspinners rotating in the CW and CCW directions, respectively, after being trapped at the beam center. These spinners rotate with their flat sides (Fig. 1e) towards the bottom of the liquid cell. Metaspinners that sedimented in a flipped orientation rotate in the opposite direction, as expected, though with lower speed and with slightly wobbling motion due to the increased and spatially varying friction against the supporting substrate.

We tracked the position and angular orientation of several metaspinners using video microscopy. For linear polarization and an incident intensity of $I \approx 75$ μW·μm$^{-2}$, corresponding to $P_0 \approx 0.87$ mW of incident power per metagrating, we found an average rotation frequency of $f = 3 \pm 0.3$ Hz (Fig. 2c). We interpret the spread in spinning performance as mainly due to slight variations in structure morphology, which can affect both the optical properties of the metagratings as well as the effective rotation drag and friction against the substrate.

The rotation equation of motion of a metaspinner can be written as $J\ddot{\varphi}(t) = \tau_{opt}(t) - \gamma_r\dot{\varphi}(t) + \tau_s(t)$, where $\varphi$ is the angle between the metagrating diffraction plane and the laboratory x-axis (Fig. 1f), $J$ is the moment of inertia, $\gamma_r$ is the rotation friction coefficient, and $\tau_{opt}$ and $\tau_s$ are the optical and the stochastic thermal torques, respectively. At the low Reynolds numbers relevant here ($Re$ ~ $10^{-4}$), inertial effects are small while Brownian diffusion is hardly noticeable. Hence, the equation of motion approximately simplifies to $\dot{\varphi}(t) = \tau_{opt}(t)/\gamma_r$.

Figure 2d shows the angular evolution $\varphi(t)$ of the RH metaspinner in Fig. 2a for four distinct polarization states. These differ drastically, with right-handed circular polarization (RCP) producing the fastest evolution, left-handed circular polarization (LCP) the lowest, and the two linear polarizations producing an intermediate and almost equal result. However, whereas $\varphi(t)$ evolves smoothly for circular polarization, linear polarization breaks the in-plane symmetry as the metaspinner rotates and therefore results in periodic oscillations that are phase-shifted by 90 deg. between horizontal (HP) and vertical (VP) polarizations. These effects are more clearly seen if data collected over several cycles is condensed in plots of angular speed versus angle $\dot{\varphi}(\varphi)$ (Fig. 2e), which according to the above is a measure of the angular variation of the optical torque.

To interpret the rotation dynamics, we developed an approximative model based on the assumption that the metagratings comprising a spinner behave as non-interacting ideal gratings excited by a plane wave (see Supporting Text for a derivation). The optical torque is obtained by adding an orbital and a spin contribution, $\tau_{opt} = \tau_{orb} + \tau_s$, where the latter is obtained by projecting the spin torque produced by the outgoing diffracted waves on the z-axis and subtracting this value from the spin torque produced by the incident field: $\tau_s = \tau_{s,in} - \tau_{s,out}$.

For HP and a RH metaspinner, we find $\tau_{orb}^{HP}(\varphi) = 2r\frac{n}{c_0}P_0\sum_i\left(T_{p,i}\cos^2(\varphi) + T_{s,i}\sin^2(\varphi)\right)\sin(\theta_i)$, where $r$ is the length of the lever arm (Fig. 1a), $P_0$ is the incident power per metagrating, $n$ the refractive index of the surrounding medium, $c_0$ the speed of light, and $T_{p,i}$ and $T_{s,i}$ are power efficiencies for p- and s-polarized diffraction into order *i* with diffraction angle $\theta_i$, respectively. In the present case, we have three transmitted and three reflected orders, but the main contribution to $\tau_{orb}^{LP}$ comes from the large difference in p-polarized transmission between order +1 and -1, since this is what the metagratings have been optimized for (Fig. 1a). The $\cos^2(\varphi)$ variation of these contributions then yield orbital torque maxima at $\varphi = 0$ and 180° for HP incidence (90 and 270° for VP), in good agreement with the experimental results.

Circularly polarized incidence is expected to produce angle-independent torques since the metaspinner then always receives the same incident power polarized parallel and perpendicular to the diffraction plane. From the grating model, we have $\tau_{orb}^{RCP} = \tau_{orb}^{LCP} = r\frac{n}{c_0}P_0\sum_i\left[\left(T_{p,i} + T_{s,i}\right)\sin(\theta_i)\right]$ for a RH spinner, which thus equals the angular average of the orbital torque for linear polarization $\langle\tau_{orb}^{HP,VP}\rangle$, while the spin torque adds or subtracts an equal amount, $\pm\tau_s$, depending on the incident light handedness. Hence, we expect that the angular average of the total optical torque for linear polarization equals the average of the total RCP and LCP torques: $\langle\tau_{opt}^{HP,VP}\rangle = \frac{1}{2}\left(\tau_{opt}^{RCP} + \tau_{opt}^{LCP}\right)$. This prediction is in excellent agreement with the data in Fig. 2d-f, which yield average rotation frequencies $\langle f\rangle$ of ~3 Hz for HP and VP, ~4.5 Hz for RCP, and ~1.5 Hz for LCP. Similarly, the rotation frequencies yield that the spin-to-orbital fraction of the optical torque for circular polarization is $\tau_s/\tau_{orb} = (f^{RCP} - f^{LCP})/(f^{RCP} + f^{LCP}) \approx 50\%$.

Figure 2g) shows simulated torques based on Maxwell's stress tensor calculations for a RH metaspinner exposed to the same incident intensity as in the experiments. The angular variations and torque ratios are clearly in excellent agreement with the experimental data. However, one notes that the simulations for HP and VP show small shifts of the peak maxima and minima that are not clearly observed there. These shifts originate in $\tau_{s,out}$, which exhibits an angular variation proportional to $\sin(2\varphi)$ but tends to zero at high diffraction angles. For the same reason, the dominant contribution to $\tau_s$ for circular polarization comes from the incident spin torque, $\tau_{s,in} = \pm\frac{2}{n\omega}P_0$.

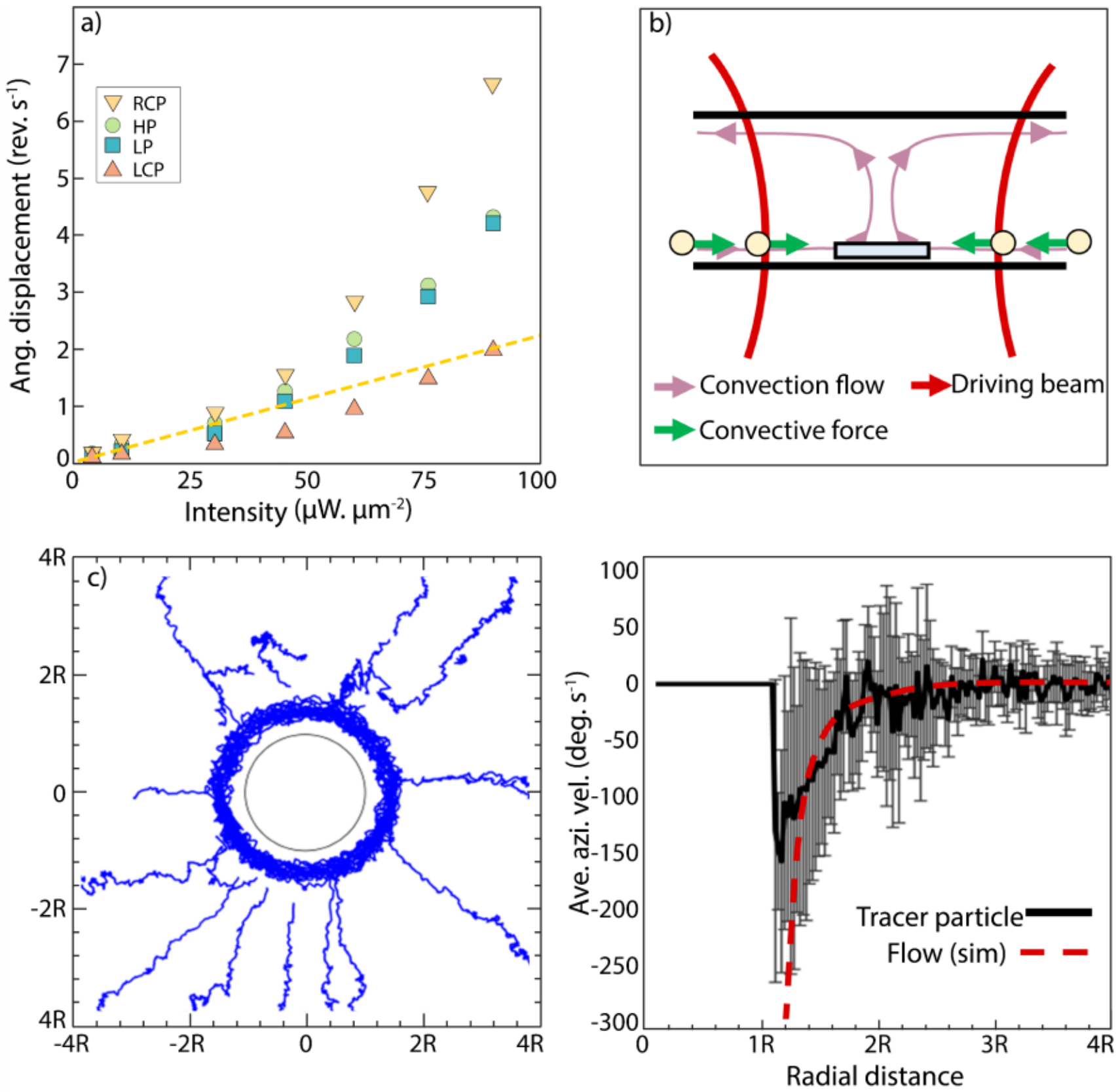

**Figure 3. Photothermal effects and flow profile.** (a) Rotation frequency of a RH metaspinner versus incident intensity for the four polarization cases. Note the supralinear variation at higher intensities. The dashed yellow line is a linear extrapolation of low intensity RCP data. (b) Schematic illustration of convective flows and forces driven by photothermal heating. (c) Blue tracks indicate tracer particle movements in the xy-plane around a metaspinner rotating at $f \approx 4.16$ Hz. The black circle indicates the metaspinner periphery. (d) Average azimuthal speed of tracer particles versus radial distance from the metaspinner c.o.m. together with simulated flow speed around a metaspinner rotating at 4.16 Hz a distance $h$ = 300 nm above a no-slip boundary.

## Photothermal effects and flow profile.

The calculated torques in Fig. 2g allow us to estimate the rotation drag coefficient by comparison to the measured rotation frequencies. For the incident intensity used (75 μW/μm$^2$), one finds $\gamma_r = \tau_{opt}/2\pi f \approx 0.5\cdot10^{-18}$ Nm·s, which is close to what is expected for a thin disk with the same radius as a metaspinner and rotating in room-temperature water ($\gamma_r = \frac{32}{3}\eta r^3 \approx 0.7\cdot10^{-18}$ Nm·s for $r = 4$ μm and viscosity $\eta = 1$ mPa·s). However, estimating $\gamma_r$ is complicated by photothermal heating, which decreases $\eta$, and by the presence of the supporting surface, which introduce hydrodynamic interactions.

We measured the rotation frequency of a RH metaspinner as a function of applied intensity to quantify the influence of photothermal heating. As shown in Fig. 3a (triangles and corresponding dashed line), $f(I)$ increases much faster than what is expected from the linear increase in optical torque with intensity. We can then estimate the temperature excess by comparing the measured $f(I)$ to a linear extrapolation from the lowest intensities. We thus assume that the supralinear variation is solely caused by a decrease in $\gamma_r \propto \eta(T)$, due to heating of the water surrounding the metaspinner, and that heating is negligible at low $I$. This gives that $f$ is a factor 2-3 higher than the interpolation at the highest intensities, indicating substantial heating and a temperature excess of the order ~40-60 K close to the metaspinner, in reasonable agreement with finite element simulations (Supporting Fig. S4).

The thermal density gradient in the water surrounding a photothermally heated metaspinner is expected to cause a convective flow, as indicated in Fig. 3b. To test this, a diluted solution of 1-µm polystyrene beads was dispersed in the sample cell and the movement of beads residing on the supporting substrate in the vicinity of a metaspinner rotating at $f \approx 4.16$ Hz ($I \approx 90$ µW·µm$^{-2}$) were tracked by video microscopy (Supporting Video 3). As summarized in Fig. 3c, the beads are indeed driven radially towards the metaspinner, as expected for convective flow, but they acquire an azimuthal speed component in the direction of the spinning motion as they get close to the metaspinner edge. However, as shown in Figure 3d, the azimuthal movement is very slow compared to the metaspinner rotation (maximum average rotation frequency ≈0.4 Hz for beads <0.3 μm from the metaspinner edge). No directed bead movement is observed unless a metaspinner is confined to the beam center (Supporting Video 4), which implies that optical gradient forces have negligible influence on the bead traces and that the convective flow is driven by photothermal heating of the metaspinner rather than of the surrounding environment.

We performed fluid dynamics simulations that included a no-slip boundary at a distance $h$ below a rotating metaspinner to interpret the azimuthal bead movement (see Methods). For realistic gap distances ($100 < h < 400$ nm)[42], the boundary is found to significantly suppress the flow compared to the case of a uniform water environment (Supporting Fig. S5). This is illustrated in Figure 3d (red dashed line), which shows that the simulated flow speed within the plane of a metaspinner rotating at 4.16 Hz at $h$ = 300 nm closely matches the measured radial variation in azimuthal bead speed and essentially vanishes a radius away from the metaspinner edge.

**Dynamics of a pair of metaspinners and the transverse lateral gradient force**

We next investigated the dynamics of pairs of metaspinners trapped side by side in the Gaussian beam. From the bead experiments (Fig. 3c), we expected that the hydrodynamic flow generated by one metaspinner would push its neighbor in the spinning direction, which in the case of two co-rotating metaspinners should result in an orbital motion in the same direction as the spinning motion. However, as summarized in Fig. 4, we observe the *opposite* behavior.

Fig. 4a-b (Supporting Video 5-6) illustrate the anomalous orbital motion for the case of pairs of LH and RH metaspinners excited with 75 $\mu W/\mu m^2$ HP illumination. The particles individually spin in the CW and CCW directions, respectively, but orbit each other in the opposite directions. The spinning frequencies are lower than for single metaspinners since the particles are now further away from the beam center, located close to the center of the orbital track, and the orbital rotation frequencies Ω are considerably lower than the spinning frequencies ($\Omega/f$ = 5-6% in Fig. 4a-b). Photothermal heating causes a supralinear increase in spinning and orbital rotation frequencies with increasing intensity but does not affect the direction of motion (Supporting Fig. S6).

To understand this effect, we first examined hydrodynamic and mechanical interactions for pairs of co-rotating metaspinners. We found that tracer beads follow the direction of fluid flow expected from the spinning direction of the individual metaspinners in a pair (Fig. 4c, Supporting Video 7). The flow around one metaspinner will thus indeed push its neighbor *in* the spinning direction, as anticipated. Hence, the hydrodynamic interaction cannot explain the anomalous orbital motion. Similarly, a mechanical interaction mechanism can be ruled out based on experimental observations. This is illustrated in Supporting Fig. S7 and Video 8, which shows two RH metaspinners that are initially orbiting in the opposite direction to their spinning motion, as in Fig. 4b, get subsequently attached to each other, and then start to orbit in the same direction as their initial spinning motion similar to two interconnected co-rotating gears. Finally, we examined pairs of counter-rotating metaspinners. In this case, the two particles also attempts to orbit in the opposite direction to their respective spinning directions, but since this is not possible due to mutual steric hindrance the pair remains essentialy stationary (Supporting Fig. S8 and Video 9).

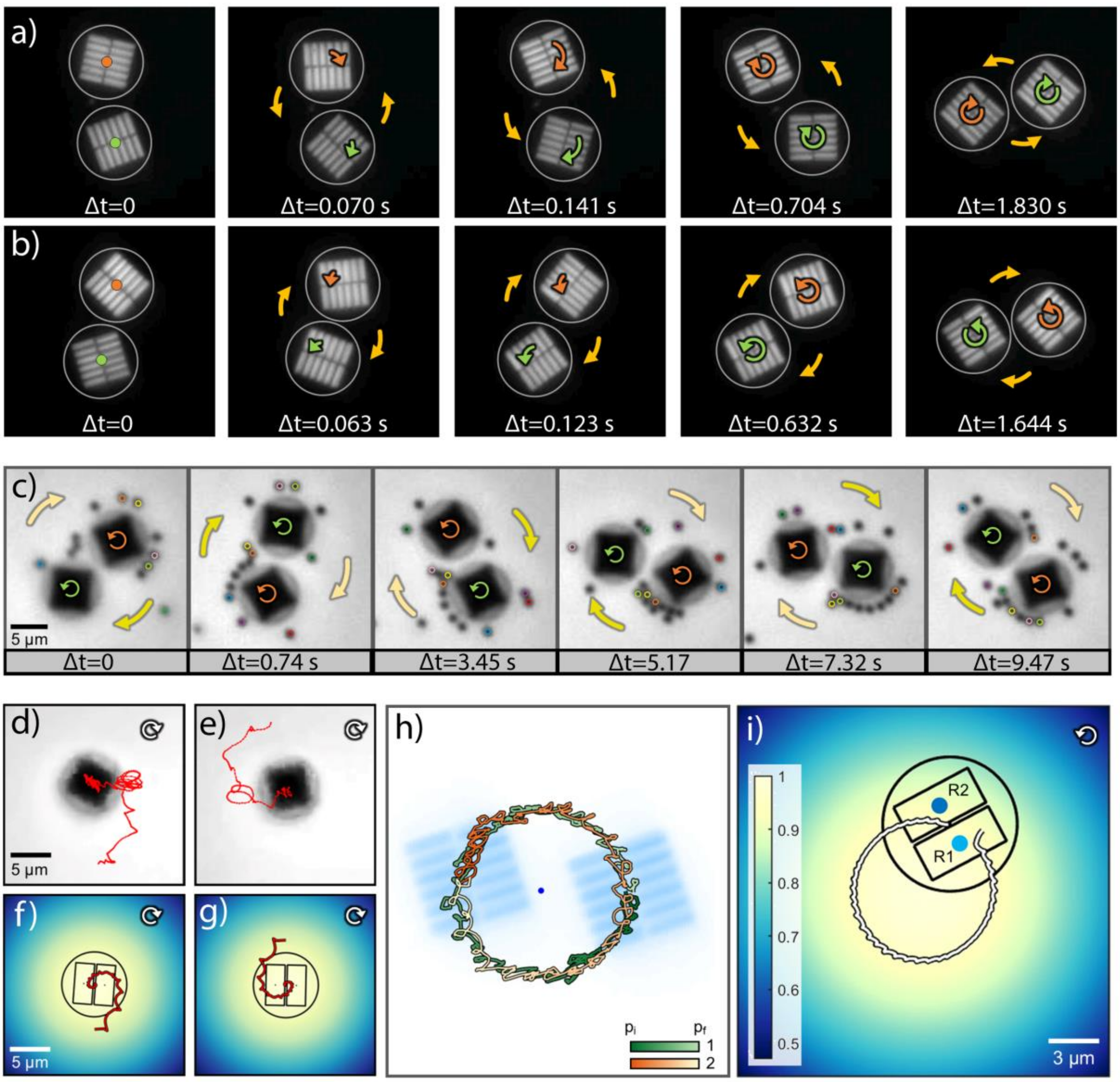

**Figure 4. Orbital dynamics of pairs of metaspinners.** (a, b) Two metaspinners rotating in the CW (a) and CCW (b) directions but orbiting each other in the opposite directions (see also Supporting Videos 5-6). (c) Movement of tracer particles around two metaspinners exhibiting simultaneous CCW rotation but CW orbiting. The particles follow the flow induced by the CCW rotation of the individual metaspinners, demonstrating that their orbital motion is not caused by hydrodynamic interactions (see also Supporting Video 7). (d, e) Tracking of single CW rotating metaspinners being pulled towards the center of the beam due to the classical optical gradient force. Note the CCW orbital motion originating in the transverse lateral gradient force. (f, g) Simulated trapping dynamics of single CW rotating metaspinners. The yellow hue indicates the extension of the Gaussian laser beam. (h) Tracking of two CCW rotating metaspinners making a full CW orbit around the beam center. The starting ($p_i$) and final ($p_f$) positions of the metaspinners are indicated as colors hues in the figure. (i) Simulated track of a single CCW rotating metaspinner that is constrained to move outside a radius of 4.1 µm from the center of the beam (see also Supporting Video 10). The initial position and orientation of metaspinner 1 in (h) were used in the model.

To elucidate the optical forces affecting the metaspinners in a pair, we first consider the classical optical gradient force, $\mathbf{F}_{grad}$, which here acts to pull a metaspinner laterally towards the center of the Gaussian beam. Experiments on $SiO_2$ disks of the same dimensions as the metaspinners but without aSi metagratings showed radial movement of the disks towards the beam center due to $\mathbf{F}_{grad}$ as expected, but no orbital movement (Supporting Fig. S9). However, since two nominally identical metaspinners in a pair cannot both occupy the beam center, they will position themselves symmetrically with their edges close to the center and with their metagratings always exposed to the Gaussian intensity gradient. Hence, the photon recoil forces generated in the metagratings comprising a metaspinner will be unbalanced and therefore not only induce an orbital torque, but also a resultant net force $\mathbf{F}_{c.o.m.}$ acting on the metaspinner center of mass (Supporting Figure S10). As outlined in the Supporting Text for a RH metaspinner subject to an intensity gradient $dI/dx$ in the $\hat{x}$-direction, one finds $\mathbf{F}_{c.o.m.} = C \cdot \{1 \cdot \hat{y} + (\sin(2\varphi)\,\hat{x} - \cos(2\varphi)\,\hat{y})\}\, dI/dx$, where $C$ is a positive definite constant determined by polarization and metagrating diffraction properties. $\mathbf{F}_{c.o.m.}$ contains two distinct components: a *transverse lateral gradient force* $\mathbf{F}_{trans}$ that always points perpendicular to the intensity gradient, and a component that varies with metaspinner orientation $\varphi$ but does not lead to a net displacement of its c.o.m. For the RH metaspinner and for $dI/dx > 0$, $\mathbf{F}_{c.o.m.}$ will lead to a continuous displacement in the positive $\hat{y}$-direction, due to $\mathbf{F}_{trans}$, and this motion will be superimposed on a circular, or in general, elliptical track due to the second component of $\mathbf{F}_{c.o.m.}$. A LH metaspinner will instead displace in the negative $\hat{y}$-direction. Hence, the direction of $\mathbf{F}_{trans}$ is determined by the metaspinner handedness, and thereby its spinning direction, and it will always be perpendicular to the intensity gradient. In a radially symmetric Gaussian intensity distribution, $\mathbf{F}_{trans}$ will instead manifest itself as an apparent azimuthal torque with respect to the beam center, forcing the metaspinner to orbit in the direction opposite to its spinning motion, as observed experimentally.

Based on the analysis above, we simulated metaspinner movements under the assumption that optical forces and torques, viscous drag, and steric hindrance dominate particle dynamics (see Methods). The simulations were performed for a linearly polarized Gaussian beam with parameters as in the experiments and utilized FDTD data to estimate the strength of the optical forces and torques, while the rotation and translation drag coefficients served as fitting parameters. As shown in Fig. 4 (d-e), a single LH metaspinner rotating clockwise and initially positioned at the periphery of the Gaussian intensity distribution is seen to gradually spiral towards the beam center while exhibiting a counterclockwise orbital movement, that is, in the opposite direction compared to its spinning motion. These effects are captured rather well in the dynamic simulations (Fig. 4 f-g) and are caused by the joint action of $\mathbf{F}_{grad}$ and $\mathbf{F}_{c.o.m.}$. The effect of steric hindrance can be approximately simulated by preventing the metaspinner c.o.m. to enter a circular area around the beam center with the same radius as the pair orbit observed experimentally. Fig 4i (Supporting Video 10) illustrates this case for a RH metaspinner representing one of the particles in the co-rotating pair shown in Fig 4h. In this case, a full clockwise orbital path is observed, as expected from the counterclockwise spinning motion of the particles.

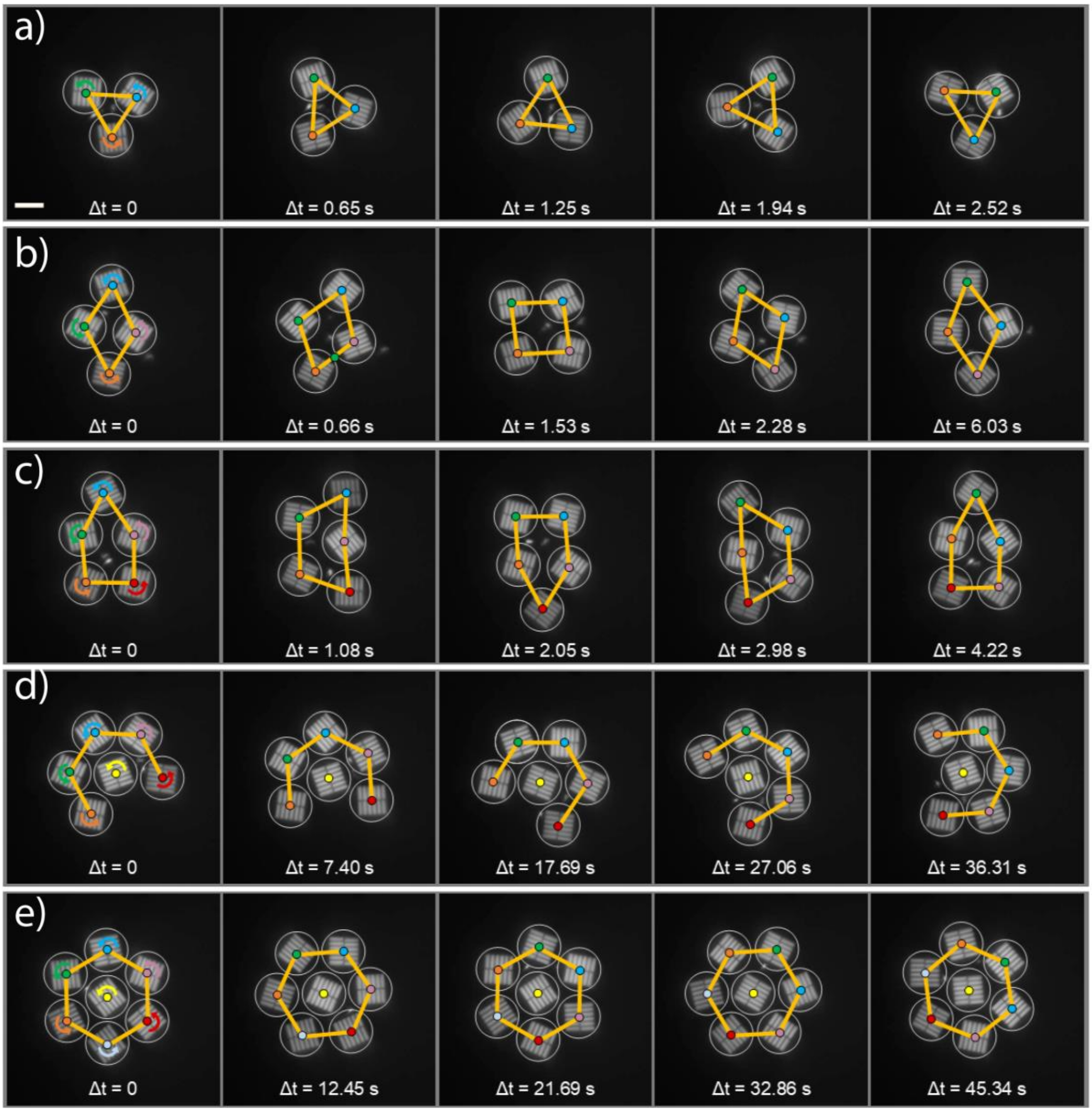

**Figure 5. Dynamic aggregates of co-rotating metaspinners.** Progressive snapshots of RH metaspinners self-assembled in a horizontally polarized Gaussian beam. Each particle spins in the counterclockwise direction but collectively orbits the beam center in the opposite direction due to the transverse lateral gradient force. The laser peak intensity is $I \approx 75$ μW·μm$^{-2}$ in (a – c) but lowered to $I \approx 30$ μW·μm$^{-2}$ in (d – e) to decrease photothermal heating effects. The beam width (with FWHM = 25.4 μm) is the same as in Fig. 2 and 4. The arrows in the first column indicate the CCW direction of rotation of each metaspinner while the colors indicate particle identity.

**Dynamic aggregates of co-rotating metaspinners**

The interplay between optical forces and steric hindrance leads to fascinating dynamic structure patterns when several metaspinners self-assemble in a Gaussian beam. Figure 5 (Supporting Video 11) shows examples of $n$ = 3-7 co-rotating RH metaspinners subject to a linearly polarized beam. While each individual metaspinner rotates in the CCW direction, with the particles closest to the beam center rotating the fastest, the aggregates again exhibit orbital motion in the opposite direction due to $\mathbf{F}_{trans}$. The structure patterns are essentially close-packed, but the particles dynamically change position within an aggregate. This is exemplified by the $n$ = 4 case, where the "blue" metaspinner initially occupies a position at the apex of the diamond shaped structure but then moves clockwise to a position along the short diagonal. An exception is when a particle manages to occupy the central position in a complete ($n$ = 7) or almost complete ($n$ = 6) hexagonal aggregate, where the gradient forces vanish and steric repulsion from the neighboring particles balances out.

## Discussion

In summary, we fabricated ultrathin untethered optically driven microscopic rotors – metaspinners – that generate torque by deflecting light in opposite directions via a moment arm through the action of two optical metasurfaces. We analyzed the optical, hydrodynamic, and photothermal properties of the structures and found that their rotational dynamics could be quantitatively explained by considering spin and orbital angular momentum transfer from the incident and deflected light together with the rotational drag caused by the surrounding liquid and supporting substrate. Moreover, we investigated pairs and larger aggregates of co-rotating metaspinners confined to a single Gaussian laser focus and found that such systems collectively orbit the optical axis in the opposite direction to the spinning motion of the individual metaspinners. This surprising effect could be explained by considering the unbalanced photon recoil forces that appear when the metasurfaces comprising a metaspinner is exposed to an intensity gradient, which results in a transverse lateral gradient force that is perpendicular to the classical optical gradient force and whose direction is determined by the metaspinner handedness.

The close-packed particle arrangements observed when several metaspinners are confined to a laser focus are similar to structures previously observed in systems of untethered microrotors coupled through hydrodynamic interactions and, as recently reported, through optical binding[46]. In the latter case, strong optical interactions between the individual nanoscopic rotors lead to phase synchronization of their spinning motions. However, no such effect is observed in the present system, that is, individual metaspinners that are located at the same radial distance from the beam center may exhibit very similar spinning frequencies but there is no clear phase correlation between their angular evolutions. This is expected since the diffracted fields emanating from the metagratings within a spinner propagate out-of-plane, implying that the in-plane optical interactions are weak in the present case.

The dynamic interaction between rotating microrotors submersed in a fluid has been the subject of intensive research directed at providing fundamental insights on the interplay between different microscale forces[1,47-49]. To the best of our knowledge, all previous studies have found that when two untethered but interacting rotors spin in the same direction, the result is an orbital motion in the same direction as the spinning motion. The opposite behavior observed here opens for interesting studies of, for example, active matter systems in which particle interactions are dominated by steric hindrances and particle trajectories are controlled by optical forces. Such studies would not necessarily have to involve optical intensity gradients since the unbalanced recoil forces behind the transverse lateral gradient force could equally well be generated by incorporating metagratings of different area or efficiency in a single metaspinner.

We note that the transverse lateral gradient force observed in this work for the case of two-dimensionally chiral metaspinners bears some resemblance to the transverse lateral optical force that can be observed when a small particle near an interface is exposed to an evanescent field[50,51]. Indeed, in the latter case, either the particle or the field needs to be chiral, such that a transverse force component is induced through spin-orbit coupling, though more work is required to establish to what extent the two effects can be explained within the same framework.

The present study utilized two oppositely directed metagratings to generate torque. Though effective, it is likely that even more efficient torque-generating structures could be found by considering metasurfaces with lower symmetry, such as spiral patterns of directional optical antennas. In the ideal case, such an optically thin structure should deflect an incident plane wave at 90 degrees in the azimuthal direction at each area element, thereby maximizing orbital angular momentum generation. However, the global optimization of such a structure, many microns in diameter, based on full Maxwell stress tensor analysis, remains a grand challenge.

Metaspinners, or similar ultrathin meatsurface-based rotors, could have interesting applications in biophysics research and applications. For example, a metaspinner, properly biofunctionalized on its flat unstructured face, could be used as a torque transducer to measure elastic and torsional properties of artificial cell membranes and, possibly, even whole cells. Furthermore, metaspinners could be used as optical stirrers or rheological sensors in microfluidic environments and devices. Here it is worthwhile to point out that the in-plane size and shape of the rotor can be easily adapted to the application needs. For example, a rotor could be equipped with protrusions able to collect small particles, such as bacteria, erythrocytes, or yeast cells, while simultaneously bringing in nutrients and dissolved gases from its photothermally induced convective flow.

Finally, optical torque-generating metasurfaces analogous to those considered here could be utilized to construct optically driven micromachinery, in which case individual rotors would need to be fixed to low friction shafts and connected through cogwheel mechanical interactions. Apart from the advantage of being able to remotely control such gear trains exclusively by light intensity and polarization, the extremely thin metasurface format implies that friction losses to viscous environments are minimized and, consequently, more energy can be spent on the intended mechanical transduction.

# Materials and Methods

*Metagrating design and characterization.* The metagratings contained within each metaspinner were optimized based on a combination of electromagnetic simulations using FDTD (Ansys Lumerical) and measurements on samples with areas large enough to determine diffraction efficiencies in transmission. Based on initial simulations and considering fabrication constraints, we settled for metagratings with three ridges and fixed periodicity ($\Lambda$ = 817 nm), corresponding to a diffraction angle in $SiO_2$ of 64 deg. The ridge widths ($W_1$, $W_2$, $W_3$), the gap distance ($g$), and the ridges height ($H$) were then optimized to yield the highest transverse momentum transfer for normally incident p-polarized plane wave illumination at $\lambda$ = 1064 nm. The simulations were based on appropriate boundary conditions, an aSi refractive index of $n_{aSi}$ = 3.8, and a uniform silica environment of $n_{SiO_2}$ = 1.45. Based on the result, a set of samples with parameters varying around the optimum were fabricated and analyzed using the setup shown in Supporting Fig. S1, resulting in the parameter set chosen for metaspinner fabrication: $W_1$ = 90 nm, $W_2$ = 130 nm, $W_3$ = 190 nm, $g$ = 100 nm, and $H$ = 490 nm. The corresponding measured power diffraction efficiencies in transmission for p- and s-polarization are shown in Supporting Fig. S2.

*Nanofabrication.* Metaspinners were fabricated on 4-inch Si wafers coated with 390 nm thermally grown $SiO_2$. A 490 nm aSi layer was deposited using low-pressure chemical vapor deposition at 550°C and the metagrating patterns were defined in an EBL process that included spin coating a positive resist (ARP6200.13), baking at 160°C for 4 minutes, exposure at 310 μC $cm^{-2}$ / 10 nA, and development for 3 min. A 60 nm Ni layer was then evaporated and lifted off to act as a hard mask. To transfer the metagrating pattern to the aSi layer, two steps of reactive-ion etching were performed, using $Cl_2$ (50 sccm HBr, 10 mtorr and 50/100 W) and HBr (40 sccm HBr, 1.5 mtorr and 340/250 W), respectively. The remaining hard mask was removed through a wet etching process. The next fabrication step defines the body of the metaspinners and involves coating the metagratings with a 610 nm thick $SiO_2$ layer using plasma-enhanced chemical vapor deposition (PECVD) and a second EBL step using the same parameters as above. After development, hard mask evaporation, and lift-off, the $SiO_2$ was etched using $CHF_3$ (12 sccm CHF3 and 17 sccm Ar, 5 mtorr, 580/50 W ICP/FW) and the residual hard mask was removed by wet etching. Finally, the metaspinners were released and dispersed in water through an 8 min. isotropic etching process using $SF_6$ gas (50 sccm SF6, 20 mtorr, 350/15 W ICP/FW). The complex refractive index of aSi was measured through ellipsometry and found to be $n \approx$ 3.8 + i0.0064 at 1064 nm.

*Simulations of optical torques.* We used the Maxwell stress tensor to calculate the torques acting on a metaspinner situated in a uniform medium with the refractive index of water ($n$ =1.33). The metaspinner consisted of two aSi metagratings, with parameters as in the optimized fabricated samples, enclosed by a cylindrical $SiO_2$ disk with diameter 8 μm and thickness 1 μm. Electromagnetic simulations based on FDTD (Ansys Lumerical) utilized perfectly matched layer boundary conditions and a source field with wavelength wavelength $\lambda_0$ =1064 nm and a Gaussian intensity distribution with beam diameter as in the experiments (FWHM=25.4 μm). Polarization was either linear or circular in the xy-plane and the metaspinner was positioned in the center of the beam.

*Simulations of flow fields and rotation drag.* We performed computational fluid dynamics simulations using the Rotating Machinery, Laminar Flow interface in COMSOL Multiphysics. The cylindrical computation volume had diameter 150 μm and height 40 μm and a metaspinner is represented by a cylinder of diameter 8 μm and thickness 1 μm located centrally at a vertical distance $h$ from the bottom interface in water. We used no-slip boundary conditions for the metaspinner surface and for the top and bottom of the computation volume, while sidewall boundaries allowed for fluid inflow and outflow. The simulations involve two mesh domains connected by a flow continuity condition: a moving inner domain containing the metaspinner and a stationary outer domain for the remaining volume. The rotation drag for a certain $d$ is obtained by setting the rotation frequency for the inner domain and calculating the resulting torque on the fluid by integrating the viscous shear stress over the domain surface. Water is considered incompressible and is characterized by dynamic viscosity $\eta = 1$ mPa·s and density $\rho = 10^3$ kg/m$^3$.

*Simulations of photothermal heating.* The temperature distribution around a metaspinner was estimated by considering the aSi gratings as the sole sources of photothermal heating and by using a complex refractive index $n_{aSi} = 3.8 + i0.0064$ obtained from ellipsometry measurements. The corresponding heat source density $Q$ was derived from simulations of absorption cross-sections obtained through FEM calculations in COMSOL. We accounted for both heat conduction and convection and used room temperature (293 K) as the boundary condition. For aSi, we utilized a thermal conductivity of 1.8 W/m·K and a density of 2329 kg/m$^3$.

*Simulations of metaspinner dynamics.* We simulated the dynamics of a metaspinner situated in a linearly polarized Gaussian beam with the same parameters as in the experiments by iteratively solving the equations of motion in the xy-plane. To simplify, we disregarded inertial effects and stochastic forces, and we also omitted the small spin contribution to the optical torque. The equations of motion then simplify to $\mathbf{F}_{opt} = \gamma_t \dot{\mathbf{r}}_{c.o.m.}$ and $\tau_{opt} = \tau_{orb} = \gamma_r \dot{\varphi}$, where $\gamma_t$ and $\gamma_r$ are the translation and rotation drag coefficients, $\dot{\mathbf{r}}_{c.o.m.}$ is the speed of the metaspinner center of mass, and $\dot{\varphi}$ is its angular speed. Based on the discussion in the Supporting Text, the optical force can be expressed as $\mathbf{F}_{opt} = \mathbf{F}_1 + \mathbf{F}_2 + \mathbf{F}_{grad}$, where $\mathbf{F}_1$ and $\mathbf{F}_2$ are the forces induced on the two metagratings in the metaspinner, while $\mathbf{F}_{grad}$ is the classical gradient force. Similarly, the orbital torque can be expressed as $\tau_{orb} = (\mathbf{r}_1 - \mathbf{r}_{c.o.m.}) \times \mathbf{F}_1 + (\mathbf{r}_2 - \mathbf{r}_{c.o.m.}) \times \mathbf{F}_2$. The local optical power needed to compute $\mathbf{F}_1$ and $\mathbf{F}_2$ were obtained by integrating the Gaussian intensity distribution over the respective metasurface area while the proportionality constant $C_{x-pol}(\varphi)$ was estimated based on comparisons with the FDTD simulations shown in main Fig. 3. Similarly, $\mathbf{F}_{grad}$ was estimated based on FDTD simulations of a $SiO_2$ disk with the same dimensions as a metaspinner (Supporting Fig. S9b). The equations of motions were solved iteratively as $\mathbf{r}_{c.o.m.}(t+\Delta t) = \Delta t \cdot \dot{\mathbf{r}}_{c.o.m.}(t) = \Delta t \cdot \mathbf{F}_{opt}(t)/\gamma_t$ and $\varphi(t+\Delta t) = \Delta t \cdot \dot{\varphi}(t) = \Delta t \cdot \tau_{orb}(t)/\gamma_r$, where the time-step $\Delta t$ was chosen small enough to result in a smooth simulated movement. The rotation and translation drag coefficients in the simulations, $\gamma_r = 5.9\times10^{-19}$ J·s and $\gamma_t = 7.3\times10^{-9}$ kg·s$^{-1}$, respectively, were used as fitting constants to obtain particle speeds similar to those observed experimentally.

## Supporting information

The online version contains Supporting material available at xxxxx.

### Acknowledgements

The authors acknowledge funding from the Knut and Alice Wallenberg Foundation (Grant No. 2016.0353 (MK), grant no. 2019.0079 (GV)). MK acknowledges scientific discussions with Dr. Alexander Stilgoe and Prof. Halina Rubinsztein-Dunlop during a stay at the University of Queensland. All nanofabrication was done at MyFab Chalmers.

### Author contributions

EE, MS, and MK conceptualized and designed the research. EE performed optical experiments, analyzed the data, and performed dynamic simulations. MS fabricated and characterized samples and performed thermal and CFD simulations. VM performed FDTD simulations. MK developed the analytical theory with the help of PJ. EE and MK wrote the paper with input from all co-authors. MK and GV supervised the study.

### Competing interests

The authors declare no competing interest.

# Supporting Information

## to

# Transverse optical gradient force in untethered rotating metaspinners

Einstom Engay[1,#], Mahdi Shanei[1,#], Vasilii Mylnikov[1], Gan Wang[2], Peter Johansson[3], Giovanni Volpe[2], and Mikael Käll[1*]

[1]Department of Physics, Chalmers University of Technology, 412 96 Gothenburg Sweden

[2]Department of Physics, University of Gothenburg, 412 96 Gothenburg, Sweden

[2]School of Science and Technology, Örebro University, 701 82 Örebro, Sweden

#Equal contributions.

*Corresponding author: mikael.kall@chalmers.se

## Supporting Figures

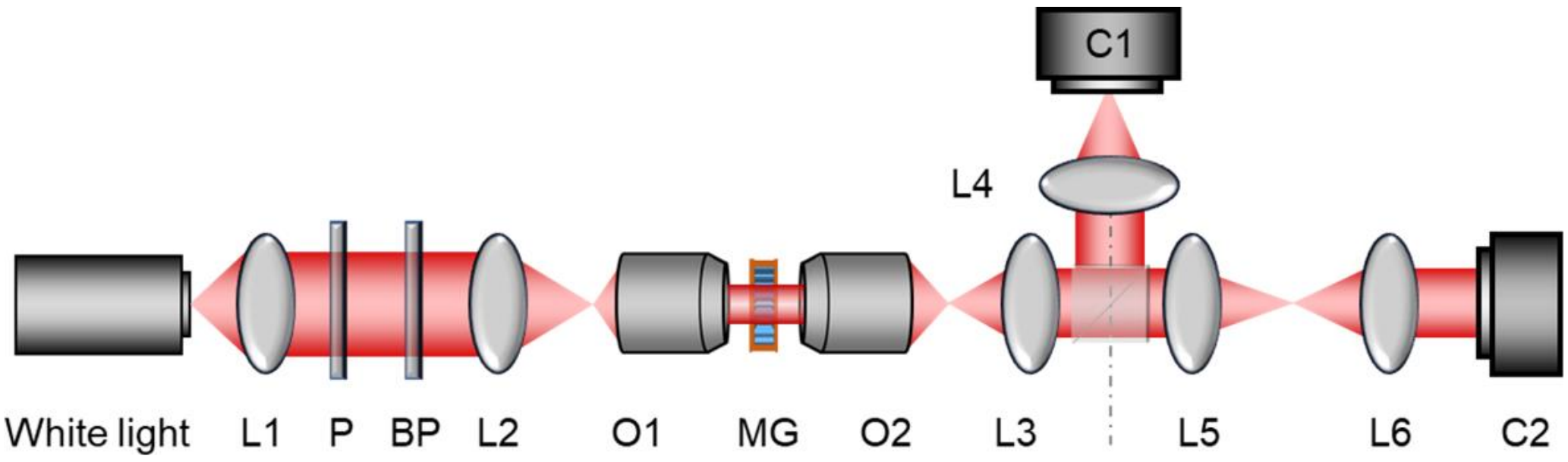

**Supporting Figure S1. Optical setup for metagrating characterization**. White light is collimated and relayed to the sample plane MG using lenses L1, L2 and microscope objective O1. The beam polarization and center wavelength are set using a polarizer, P and a bandpass filter, BP, with transmission window $\lambda = 1064 \pm 10$ nm. The metagrating is imaged onto camera 2 (C2, Thorlabs Zelux) using a 100x oil-immersion objective O2 with NA = 1.49 (Nikon) and a series of relay lenses (L3, L5 and L6). A beam splitter, a Fourier transforming lens (L4) and second camera 1 (C1, Thorlabs Zelux) are employed to record the diffraction pattern in the Fourier/*k*-space.

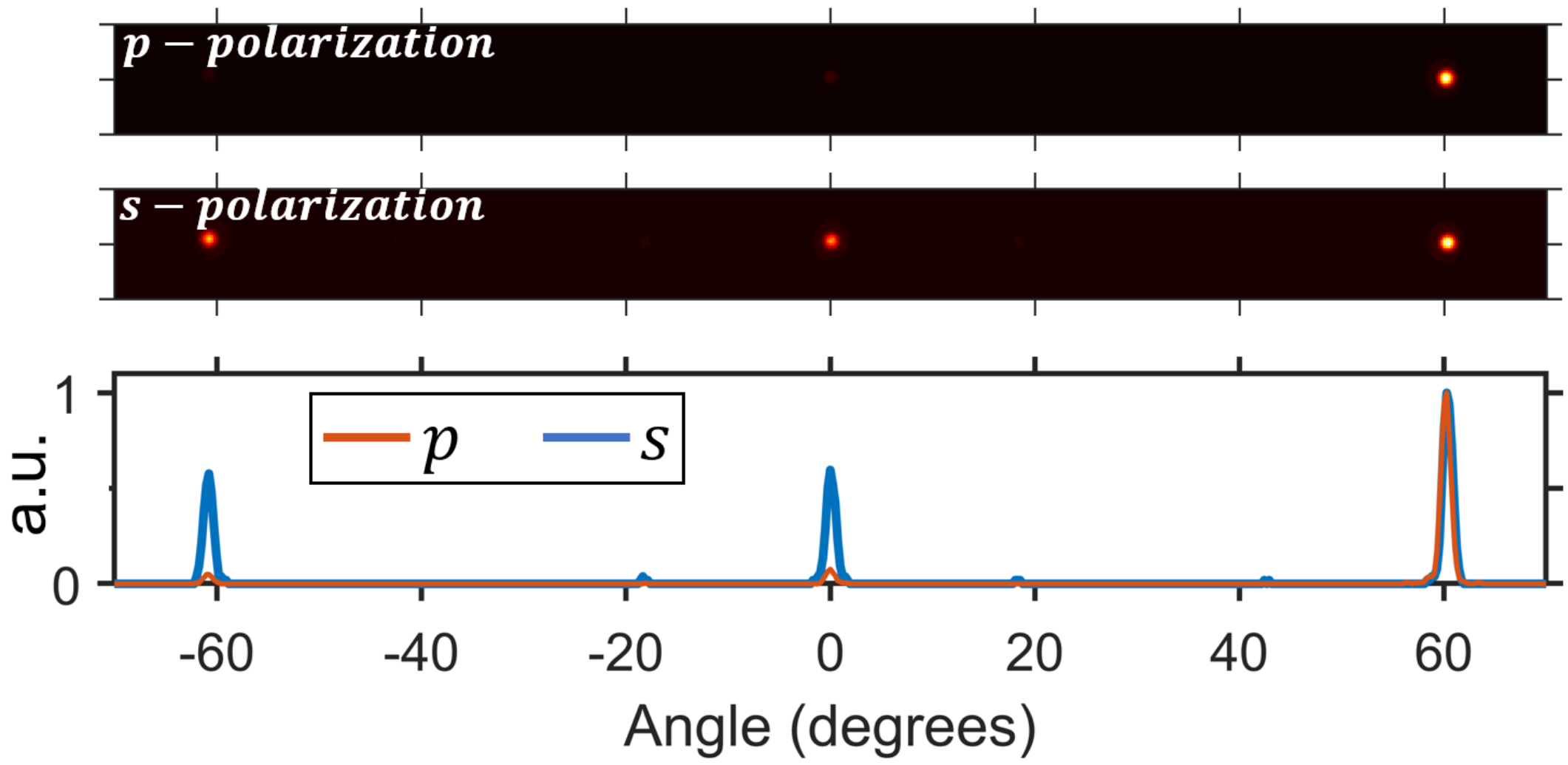

**Supporting Figure S2. Measured diffraction efficiencies in transmission.** The diffraction efficiencies of large area (~0.1x0.1 $mm^2$) metagratings were recorded in transmission using Fourier imaging (Figure S1). For the optimized structure, the overall transmission coefficients for p- and s-polarization were found to be 70% and 44%, respectively. The power diffraction efficiencies into the three allowed transmission orders were $T_{p,+1}^{exp} = 63$ %, $T_{p,0}^{exp} = 3.8$ %, and $T_{p,-1}^{exp} = 2.7$ % for p-polarization and $T_{s,+1}^{exp} = 19.3$ %, $T_{s,0}^{exp} = 11.4$ %, and $T_{s,-1}^{exp} = 12.5$ % for s-polarization. The experimental data agree reasonably well with FDTD results (overall transmission coefficient 75%, $T_{p,+1}^{FDTD} = 69$%, $T_{p,0}^{FDTD} = 8.6$ %, $T_{s,-1}^{FDTD} = 5.8$ % for p-polarization, and 70%, $T_{s,+1}^{FDTD} = 35.6$%, $T_{s,0}^{FDTD} = 15.5$ %, and $T_{s,-1}^{exp} = 21.4$ % for s-polarization).

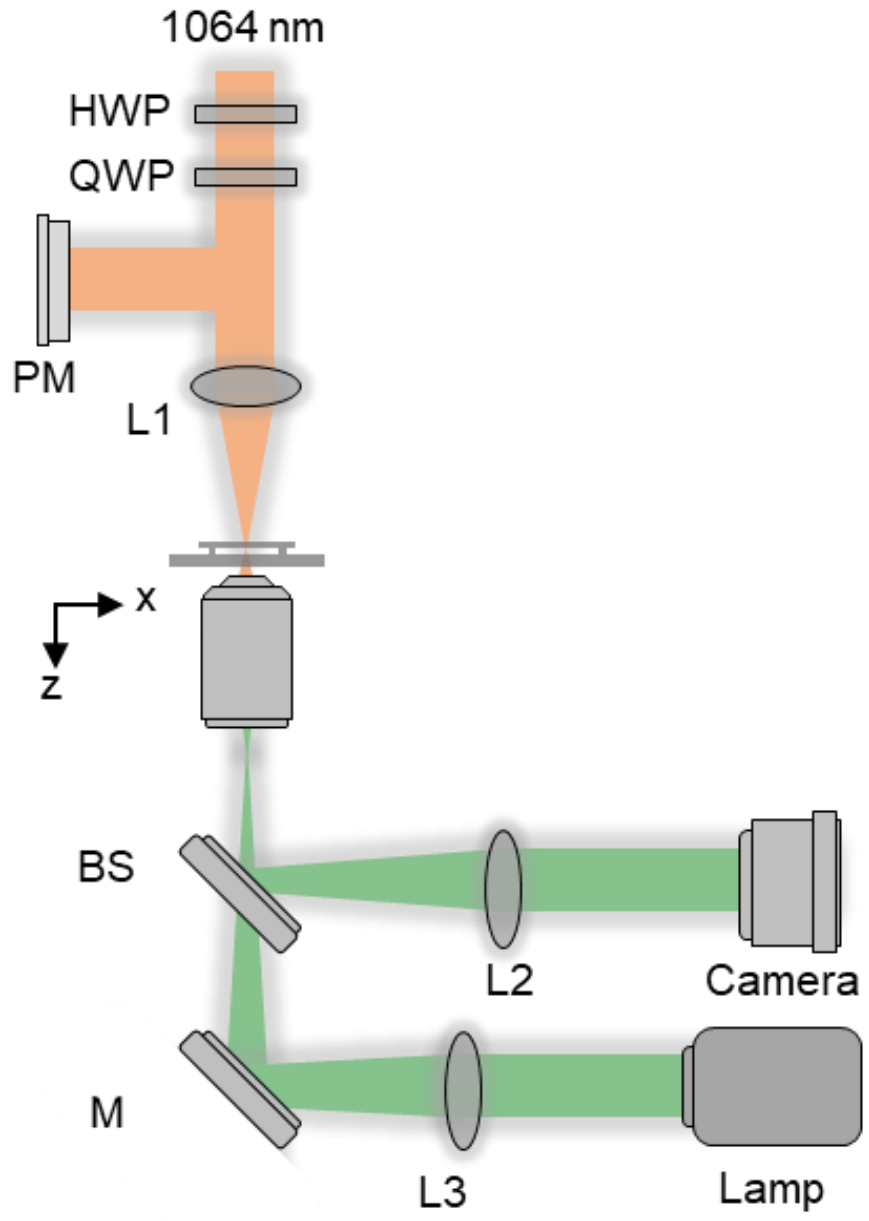

**Supporting Figure S3. Setup for optical rotation experiments.** Measurements were performed in an inverted microscope (Nikon Eclipse Ti) coupled to a λ = 1064 nm CW laser (Cobolt Rumba 2W) using suitable polarization and focusing optics. The metaspinner sample is contained in a thin liquid cell on an automated stage and laser illuminated from above using a *f* = 5 cm lens (L1). The sample is observed in reflection from below using a dry objectives (40X, NA = 0.95 or 20X, NA = 0.7), white-light illumination and a CMOS camera (Thorlabs Zelux). HWP, half-wave plate; QWP, quarter-wave plate; PM, power meter; L, lens; BS, beam splitter; M, mirror.

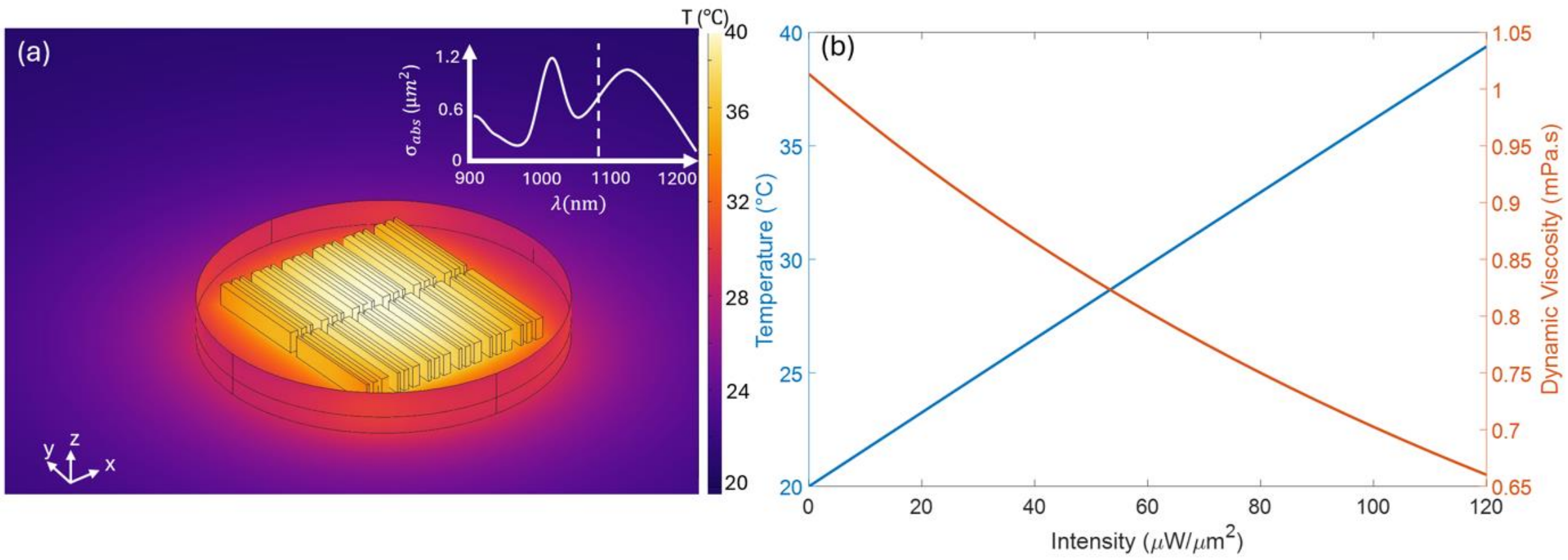

**Supporting Figure S4. Finite element simulation of temperature distribution around a metaspinner.** (a) FEM simulation of absolute temperature around a metaspinner illuminated by a 100 µW/µm² s-polarized incident plane wave with wavelength λ = 1064 nm. The calculated heat flux density in the metagratings is based on an aSi refractive index of n = 3.8 + i0.0064 at 1064 nm. The inset shows the calculated absorption spectrum of a metagrating unit cell with width 2.5 µm. (b) Simulated maximum temperature around the metaspinner as a function of incident light intensity and the corresponding calculated dynamic viscosity of water.

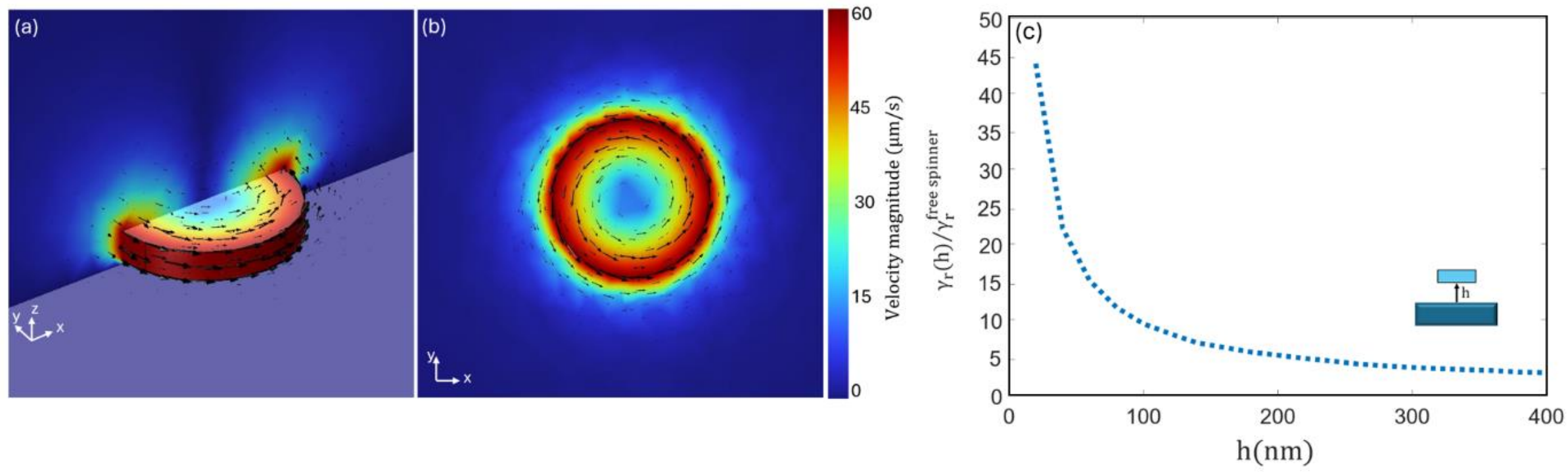

**Supporting Figure S5. Fluid dynamics simulations of flow field around a rotating metaspinner.** (a)-(b) simulation of the velocity pattern of water around a metaspinner rotating at an angular velocity of 15 rad/s. The spinner is positioned at a surface-separation of 300 nm from the bottom no-slip boundary of the calculation domain. The upper no-slip boundary is located 50 μm from the bottom, while the domain sidewall boundaries are free, permitting fluid inflow and outflow. The simulations involve two distinct mesh domains: a moving domain containing the spinner and a stationary domain encompassing the rest of the model. A flow continuity condition was applied to the boundary between these domains. c) fluid dynamics simulation of metaspinner rotation drag coefficient $\gamma_r(h)$ versus distance $h$ from bottom surface in units of $\gamma_r = 7.5\cdot10^{-19}$ Nm·s for a “free” metaspinner located in the middle of the simulation domain.

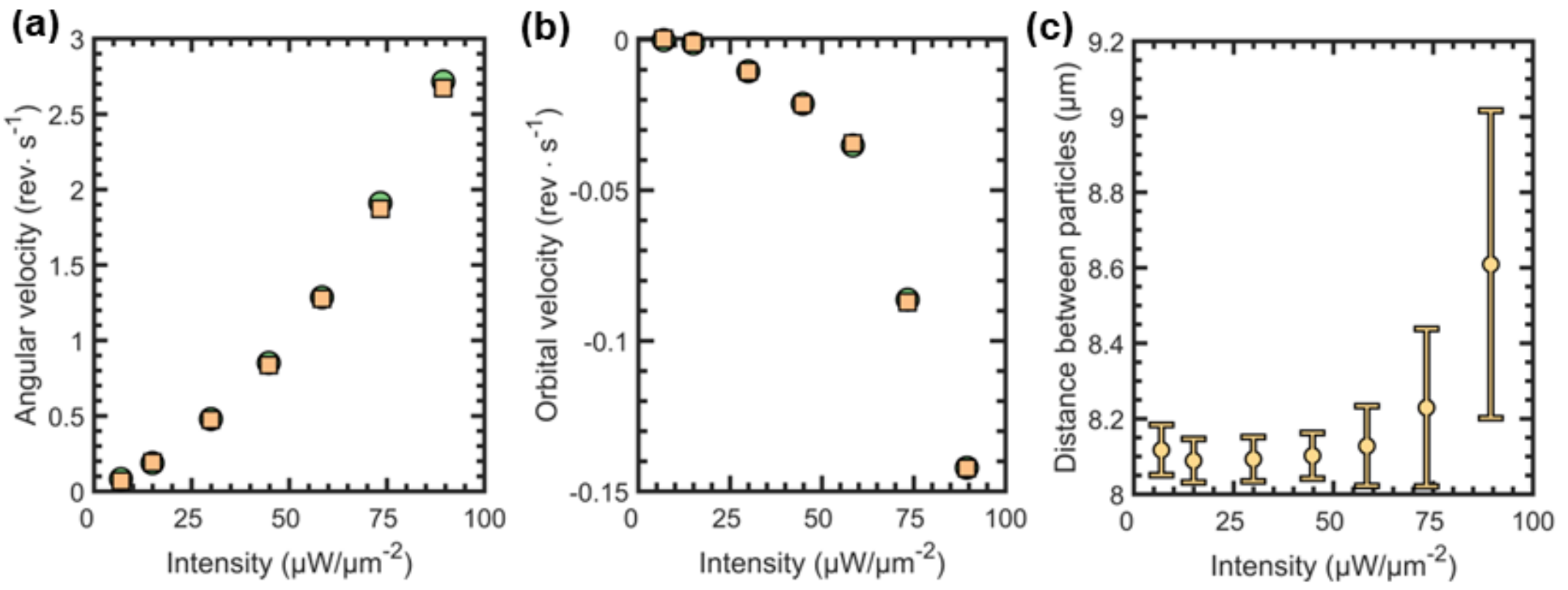

**Supporting Figure S6: Orbital rotation versus applied intensity.** (a) Spinning frequencies versus intensity for a pair of co-rotating metaspinners. (b) Orbital frequency of the pair versus applied intensity. (c) Estimated center-to-center separation distance between the two metaspinners as a function of applied intensity. The “error-bars” indicate the spread in distances observed over several full trajectories.

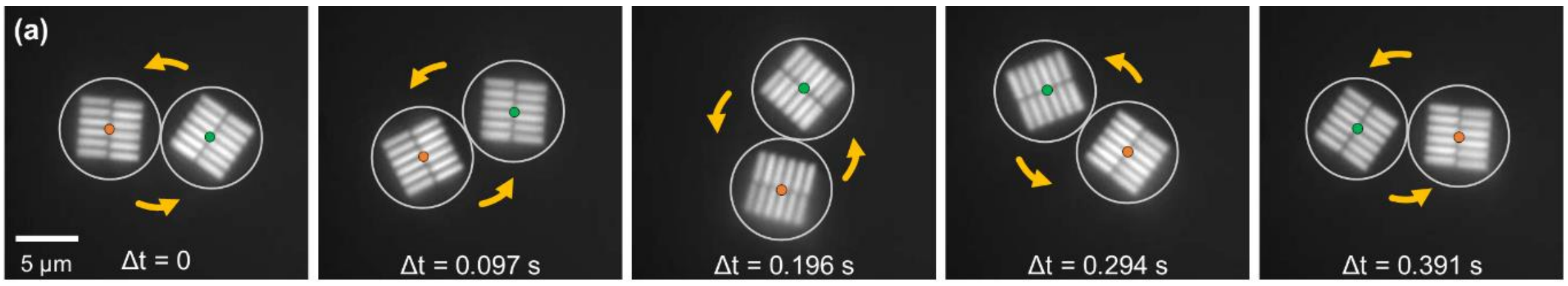

**Supporting Figure S7: Effect of mechanical locking.** (a) A pair of metaspinners that are initially separate, co-rotating in CCW about their own axes and orbiting in CW, begin to orbit in the opposite direction (CCW) when they accidentally lock together (see also Supplementary Video 8).

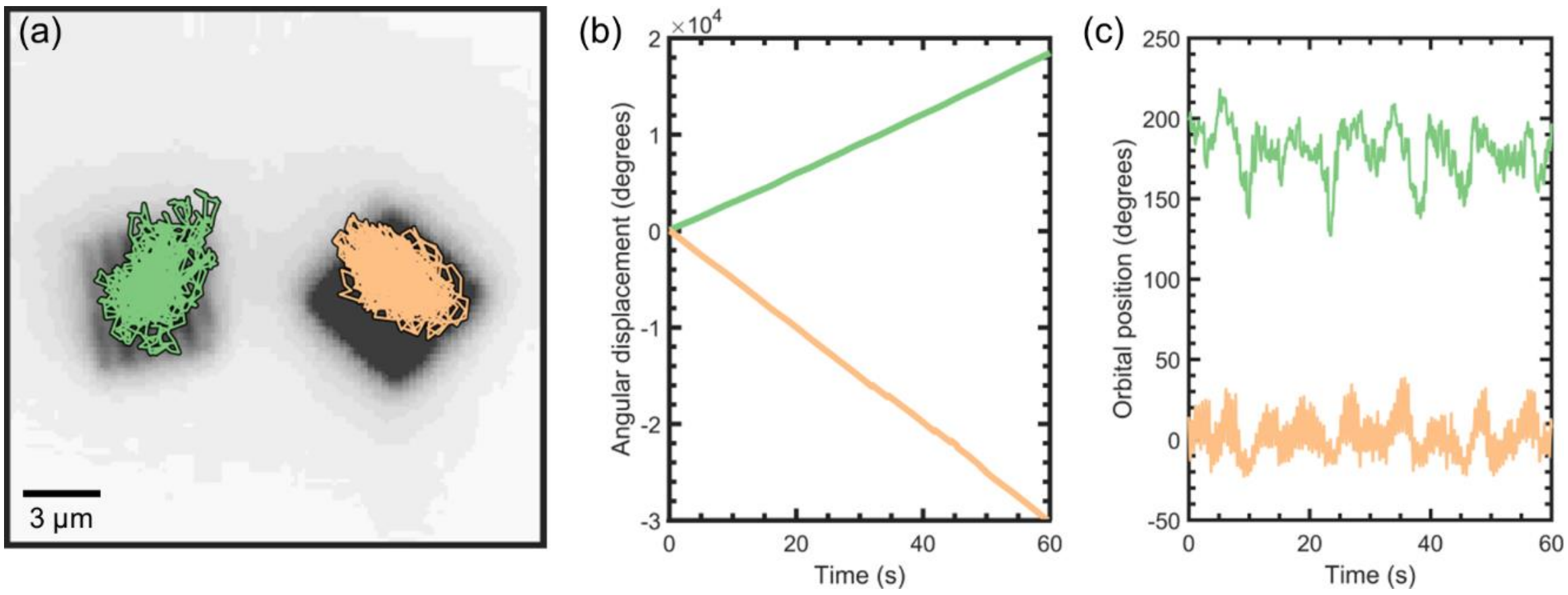

**Supporting Figure S8**: **Dynamics of a pair of counter-rotating metaspinners.** (a) Tracks of the centroids of a pair of counter-rotating metaspinners subject to a horizontally polarized beam with intensity 60 μW/μm$^2$ and the same width as in main Fig. 4. (b) Angular displacement of the metaspinners as they rotate. (c) Orbital position of each metaspinner. The two metaspinners have slightly different spinning frequencies ($f_1 \approx 0.9$ Hz, $f_2 \approx 1.4$ Hz) but are rotating in opposite directions. The metaspinners still exhibit translational movements consistent with the generated transverse optical gradient force. However, unlike pairs of co-rotating metaspinners, the counter-rotating pair does not exhibit any consistent orbital rotation because of the mutual steric hindrance between the two spinners.

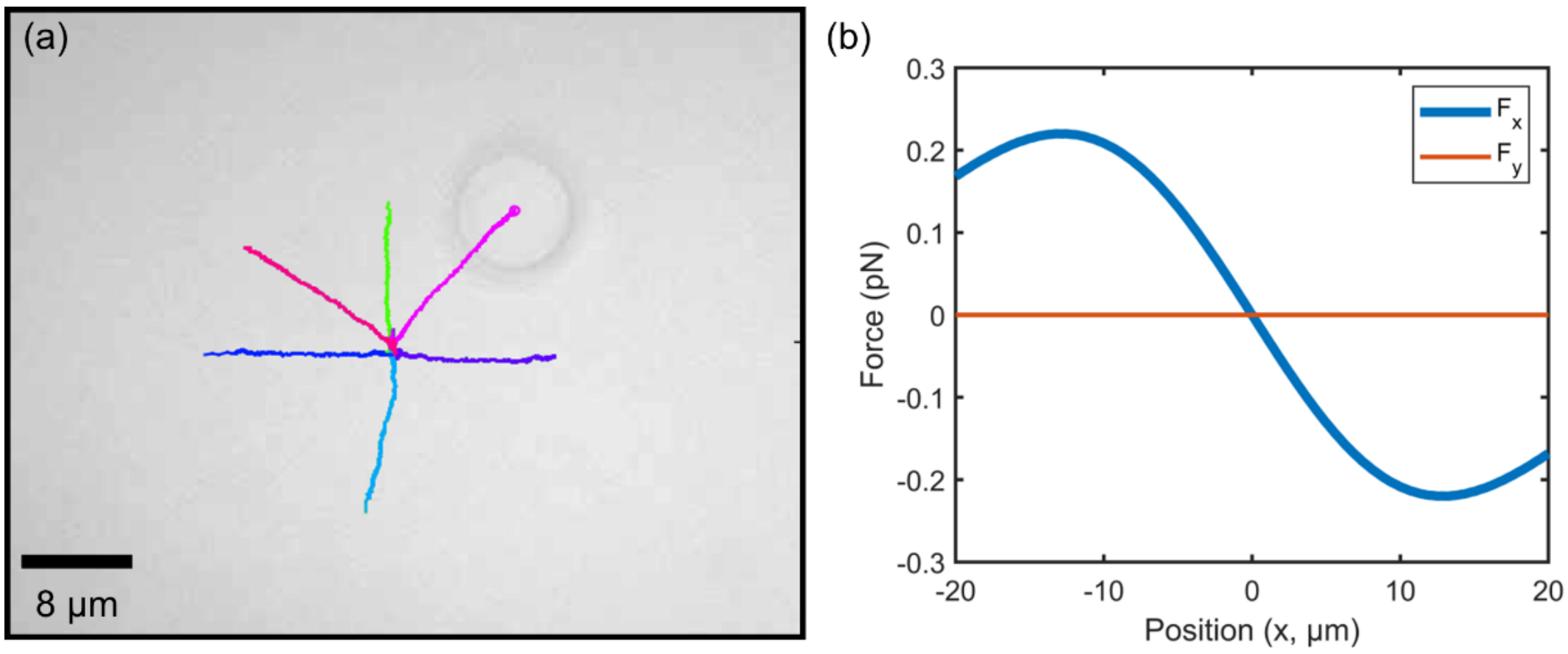

**Supporting Figure S9: Classical gradient force acting on a $SiO_2$ disk.** (a) Tracks of a $SiO_2$ disk subject to a x-polarized Gaussian beam with the same width as in main Fig. 4 and intensity 90 µW/µm$^2$. The track colors indicate different starting positions. The SiO2 disk has the same dimensions as a metaspinner but it moves radially towards the beam center without any sign of spiral or orbital motion. (b) Lateral force components $F_x$ and $F_y$ versus $SiO_2$ disk displacement from the center of a Gaussian beam with the same characteristics as in a) based on FDTD Maxwell stress tensor analysis.

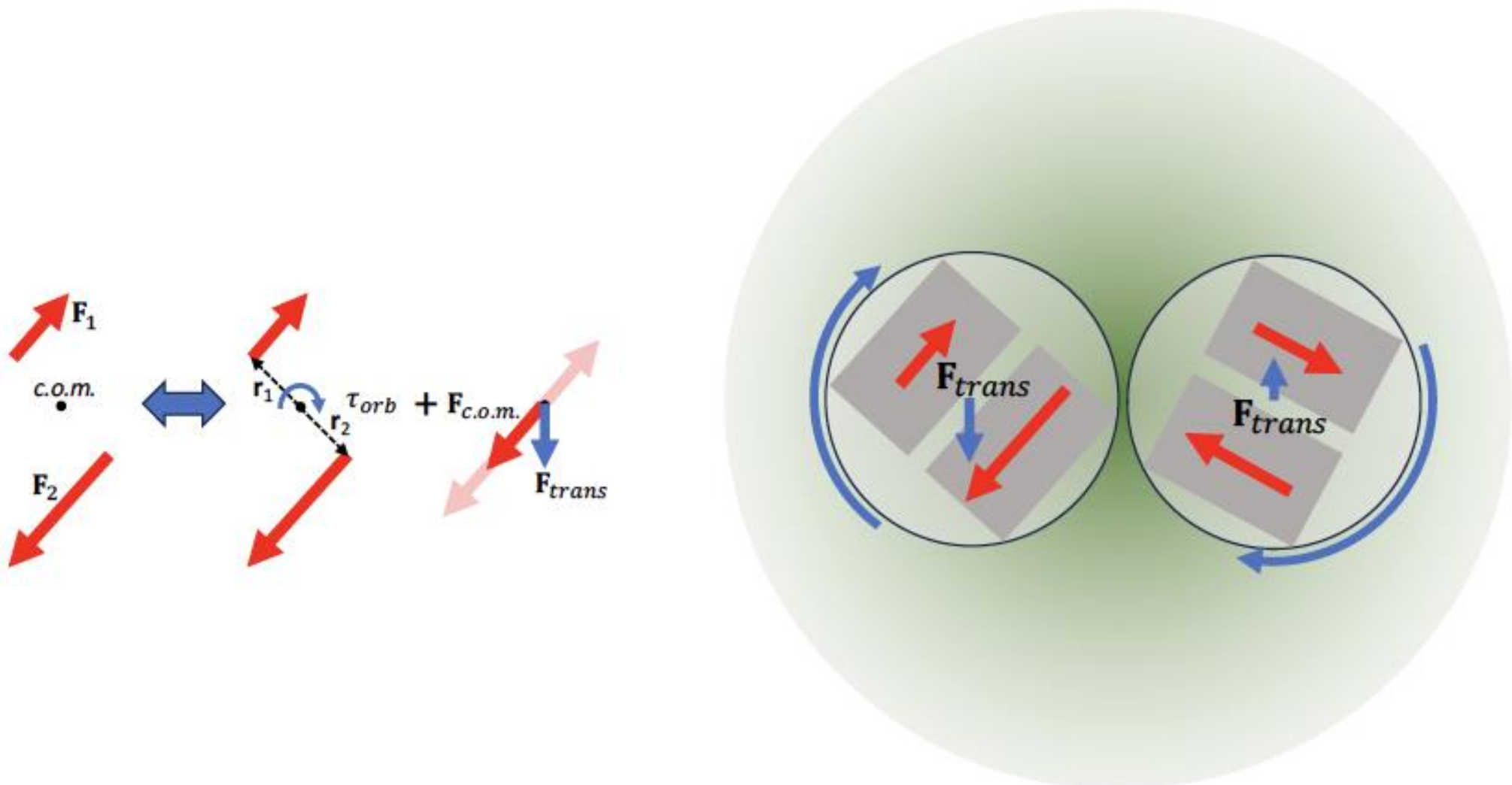

**Supporting Figure S10**: **Schematic illustration of optical forces and torques acting on a pair of metaspinners in a Gaussian beam.** The photon recoil forces $\mathbf{F}_1$ and $\mathbf{F}_2$ acting on the two metagratings in a left-handed (LH) metaspinner situated in a Gaussian intensity gradient result in an orbital torque, $\tau_{orb}$, which spins the particle around its own axis in the clockwise direction, and a force, $\mathbf{F}_{c.o.m.}$, acting on its center of mass. The force component that is transverse to the radial intensity gradient, $\mathbf{F}_{trans}$, causes the spinner to orbit around beam center in the counterclockwise direction, that is, opposite to the spinning motion. Metaspinner movement towards the beam center is driven by the classical optical gradient force (not shown) but blocked through steric hindrance from the other particle in the pair. Hydrodynamic interaction forces work against $\mathbf{F}_{trans}$ but are weak compared to the optical forces.

# Supporting Videos

**Supporting Video 1.** LH metaspinner rotating in the CW direction. The metaspinner is illuminated by a HP incident beam with peak intensity $I \approx 75\ \mu W \cdot \mu m^{-2}$ (actual frame rate).

**Supporting Video 2.** RH metaspinner rotating in the CCW direction. The metaspinner is illuminated by a HP incident beam with peak intensity $I \approx 75\ \mu W \cdot \mu m^{-2}$ (actual frame rate).

**Supporting Video 3.** A single RH metaspinner with tracer particles under HP illumination with peak intensity $I \approx 90\ \mu W \cdot \mu m^{-2}$ (actual frame rate).

**Supporting Video 4.** Tracer particles under HP illumination with peak intensity $I \approx 90\ \mu W \cdot \mu m^{-2}$ (actual frame rate).

**Supporting Video 5.** A pair of LH metaspinners co-rotating in the CW direction while mutually orbiting in the CCW direction. The pair is illuminated by a HP incident beam with peak intensity $I \approx 75\ \mu W \cdot \mu m^{-2}$ (actual frame rate).

**Supporting Video 6.** A pair of RH metaspinners co-rotating in the CCW direction while mutually orbiting in the CW direction. The pair is illuminated by a HP incident beam with peak intensity $I \approx 75\ \mu W \cdot \mu m^{-2}$ (actual frame rate).

**Supporting Video 7.** a) RH metaspinner dimer with tracer particles under HP illumination with peak intensity $I \approx 90\ \mu W \cdot \mu m^{-2}$ (1/3 actual frame rate). b) RH metaspinner dimer with tracer particles under HP illumination with peak intensity $I \approx 90\ \mu W \cdot \mu m^{-2}$ (actual frame rate).

**Supporting Video 8.** a) A pair of RH metaspinners initially co-rotating in the CCW direction while mutually orbiting in the CW direction, which later get attached to each other and collapse into one unit that rotate in the CCW direction. The pair is illuminated by a HP incident beam with peak intensity $I \approx 45\ \mu W \cdot \mu m^{-2}$. b) A collapsed pair of RH metaspinners illuminated by a HP incident beam with peak intensity $I \approx 90\ \mu W \cdot \mu m^{-2}$ (actual frame rates).

**Supporting Video 9.** A pair of counterrotating metaspinners. The pair is illuminated by a HP incident beam with peak intensity $I \approx 60\ \mu W \cdot \mu m^{-2}$ (actual frame rate).

**Supporting Video 10.** Dynamic simulation of a RH metaspinner restricted to move within 4.1 μm from the beam center for HP illumination with peak intensity $I \approx 75\ \mu W \cdot \mu m^{-2}$ (actual frame rate).

**Supporting Videos 11.** Aggregates of $N$ = 3-7 co-rotating RH metaspinners. HP illumination with peak intensity $I \approx 75\ \mu W \cdot \mu m^{-2}$ for $N$ = 3-5 and intensity $I \approx 30\ \mu W \cdot \mu m^{-2}$ for $N$ = 6 – 7 (actual frame rates).

# Supporting Text

In the following, we derive simplified equations that can be used to interpret the experimental results discussed in the main paper. We neglect edge diffraction and assume that the gratings comprising a spinner are irradiated by plane waves incident normal to the grating surface. We only consider forces and torques within the plane of the grating, since this is what we observe experimentally, and we neglect absorption.

**Incident and diffracted fields**

We define a metagrating coordinate system (x', y', z') such that the plane of the grating lies in x'y'-plane and diffraction occurs in the x'z'-plane. The diffraction angles $\theta_i$, where $i$ signifies a diffraction order, are measured from the z'-axis and transmitted and reflected orders are treated on an equal footing. The metagrating is situated in a homogeneous medium with refractive index $n$. The torques and transverse forces are obtained from projections on the z'-axis and the x'y'-plane, respectively.

The incident polarization and the forces and torques on the grating are defined/measured in a laboratory frame (x, y, z) with z//z´. We focus on two polarization cases:

(1) $\mathbf{E}_{xy}^{0} = E_0 Re\left\{\begin{pmatrix}1\\0\end{pmatrix} e^{i(kz-\omega t)}\right\}$, x-polarized incidence.

(2) $\mathbf{E}_{xy}^{0} = {E_0}/{\sqrt{2}}\, Re\left\{\begin{pmatrix}1\\i\end{pmatrix} e^{i(kz-\omega t)}\right\}$, LCP incidence.

Here $Re$ means taking the real part, $E_0$ is the amplitude of the incident electric field with wavenumber $k$ and angular frequency $\omega$. Assume that the grating axis x' forms an angle $\varphi$ relative to x. To work out the forces and torques, we need to translate from the grating to the laboratory frame by applying the rotation matrix:

(3) $\mathbf{E}_{x'y'}^{0} = \begin{pmatrix}\cos(\varphi) & \sin(\varphi)\\ -\sin(\varphi) & \cos(\varphi)\end{pmatrix} \mathbf{E}_{xy}^{0}$.

The amplitudes and phases of the diffracted fields are defined through complex 2x2 tensors in planes perpendicular to the propagation directions of the diffracted waves. The tensors will be diagonal since an incident field polarized along x' will exclusively generate p-polarized diffraction while an incident field polarized along y' will generate s-polarized diffraction. Diffraction order *i* is thus characterized by the tensor

(4) $\begin{pmatrix}p_i & 0\\ 0 & s_i e^{i\psi^i}\end{pmatrix}$,

where $p_i$ and $s_i$ are real numbers (amplitudes) and $\psi^i$ is the relative phase shift between the p- and s-polarized components. A metagrating with three transmission and three reflection orders, as is the case considered here, is thus described by 12 amplitudes and 6 relative phases. Multiplying the diffraction tensor with the incident field in the (x', y') frame, and omitting $Re$ and the propagation phase for brevity, yields the diffracted fields in the tilted ($p_i$, $s_i$) frames as:

(5) $\mathbf{E}_{ps}^{i} = E_0 \begin{pmatrix} p_i \cos(\varphi) \\ -s_i e^{i\psi^i} \sin(\varphi) \end{pmatrix}$ for x-polarized incidence, and

(6) $\mathbf{E}_{ps}^{i} = {E_0}/{\sqrt{2}} \begin{pmatrix} p_i[\cos(\varphi) + i\sin(\varphi)] \\ s_i e^{i\psi^i}[-\sin(\varphi) + i\cos(\varphi)] \end{pmatrix}$ for LCP incidence.

The corresponding field amplitudes squared are:

(7) ${E_{ps}^{i}}^2 = {E_0}^2\{p_i^2 \cos^2(\varphi) + s_i^2 \sin^2(\varphi)\}$ for x-polarized incidence, and

(8) ${E_{ps}^{i}}^2 = \frac{{E_0}^2}{2}\{p_i^2 + s_i^2\}$ for LCP incidence.

**Reaction force acting on a single grating.**

The Minkovski linear momentum $\wp$ carried by a beam with power $P$ per time unit is

(9) $\frac{\partial \wp}{\partial t} = n\frac{P}{c_0}$,

where $n$ is the refractive index. By projecting the diffracted momenta on the x'-axis, through multiplication with $\sin(\theta_i)$, correcting for the change in wave-front area from the incident to the diffracted beams through multiplication with $|\cos(\theta_i)|$, to ensure energy conservation, we obtain the net reaction force in the x'-direction by summing over all diffraction orders and taking the negative. For x-polarized incidence, we get:

(10) $F_{x'} = -\frac{n}{c_0} P_0 \sum_i \left[p_i^2 \cos^2(\varphi) + s_i^2 \sin^2(\varphi)\right] \cdot \sin(\theta_i) \cdot |\cos(\theta_i)| = -\frac{n}{c_0} P_0 \sum_i \left[T_{p,i} \cos^2(\varphi) + T_{s,i} \sin^2(\varphi)\right] \cdot \sin(\theta_i)$

while for LCP, we get:

(11) $F_{x'} = -\frac{n}{2c_0} P_0 \sum_i \left[p_i^2 + s_i^2\right] \cdot \sin(\theta_i) \cdot |\cos(\theta_i)| = -\frac{n}{2c_0} P_0 \sum_i \left[T_{p,i} + T_{s,i}\right] \cdot \sin(\theta_i)$.

We have here utilized that the incident power $P_0$ that hits the grating is given by the incident irradiance (intensity) $I_0$ times the grating area $A$ as $P_0 = I_0 A$, while the irradiance is given by the incident electric field amplitude $E_0$ as $I_0 = \frac{\epsilon_0 c}{2} E_0^2$, and we have introduced power transmission/reflection coefficients $T_{p,i} = p_i^2 |\cos(\theta_i)|$ and $T_{s,i} = s_i^2 |\cos(\theta_i)|$ for the different polarizations and diffraction orders.

Finally, since there is no force in the y'-direction, the force components in the (x,y) system simply becomes:

(12) $$\mathbf{F}_{xy} = \begin{bmatrix} \cos(\varphi) F_{x'} \\ \sin(\varphi) F_{x'} \end{bmatrix}$$

**Spin torque acting on a single grating.**

The spin density of a plane wave is directed parallel to the propagation direction and given by:

(13) $$\langle \mathbf{s} \rangle = Im[\epsilon_0 \mathbf{E}^* \times \mathbf{E}]/2\omega$$

By inserting the incident and diffracted fields above, we can calculate the spin densities for the two polarization cases considered. After some algebra, one obtains for x-polarized incidence:

(14) $$\langle \mathbf{s}_0 \rangle = 0,$$

(15) $$\langle \mathbf{s}^i \rangle = -\hat{k}_i \frac{\epsilon_0 E_0^2}{2\omega} p_i s_i \sin(\psi_i) \sin(2\varphi),$$

while LCP incidence gives:

(16) $$\langle \mathbf{s}_0 \rangle = \hat{z} \frac{\epsilon_0 E_0^2}{2\omega},$$

(17) $$\langle \mathbf{s}^i \rangle = \hat{k}_i \frac{\epsilon_0 E_0^2}{2\omega} p_i s_i \cos(\psi_i),$$

where $\hat{k}_i$ indicates the direction of propagation of the diffracted waves.

We can now write down the torque exerted on a grating with area $A$ by following the same procedure as for the linear momentum transfer and forces in the xy-plane:

(18) $$\tau_{\hat{z}} = \frac{c_0}{n} A \langle \hat{z} \cdot \mathbf{s}_0 \rangle - \frac{c_0}{n} A \sum_i \langle \hat{k}_i \cdot \mathbf{s}^i \rangle \cos(\theta_i) \, |\cos(\theta_i)|,$$

where the first $\cos(\theta_i)$ factors project out the actual $\hat{z}$-components of the diffracted torque. Writing out the full expression for the torques in terms of power, we obtain for x-polarized incidence:

(19) $$\tau_{\hat{z}} = \frac{1}{n\omega} P_0 \sin(2\varphi) \sum_i p_i \, s_i \sin(\psi_i) \cos(\theta_i) \, |\cos(\theta_i)|,$$

while for LCP, we find:

(20) $$\tau_{\hat{z}} = -\frac{1}{n\omega} P_0 \{1 - \sum_i p_i \, s_i \cos(\psi_i) \cos(\theta_i) |\cos(\theta_i)|\}.$$

The torque obtained for x-polarization can be thought of as an "alignment torque" since it strives to orient the grating parallel to the incident polarization due to the $\sin(2\varphi)$ factor. The magnitude of the alignment torque is maximum for $\varphi = \pm 45^{\circ}$ and $\pm 135^{\circ}$ and it vanishes when the grating is either aligned with or perpendicular to the polarization. The ”spin torque” obtained for LCP incidence is, in contrast, independent of grating orientation.

**Torques acting on a metaspinner**

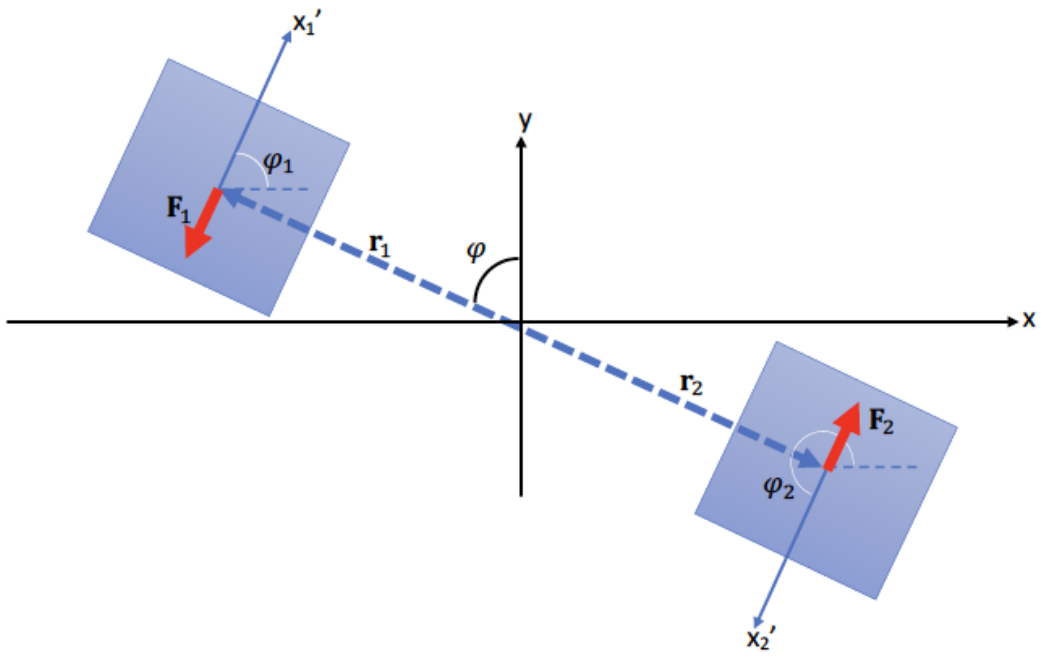

We describe a metaspinner as two gratings connected by a rigid bar, as in the figure above. The gratings are oriented such that the xy-plane projections of their main diffraction orders are perpendicular to the bar and points in opposite directions. We can then add the torques from each grating to obtain the total torque on the spinner. The torques originating in the translation forces are classified as “orbital torque” while the torques derived from polarization conversion are classified as “spin torque”. The figure above illustrates a right-handed (RH) metaspinner, designed to rotate in the counterclockwise (CCW) direction. For this case, we have

$$\varphi_1 = \varphi, \varphi_2 = \varphi + \pi,$$

(21) $$\mathbf{r}_1 = r\begin{bmatrix}-\sin(\varphi)\\ \cos(\varphi)\end{bmatrix}, \mathbf{r}_2 = r\begin{bmatrix}-\sin(\varphi+\pi)\\ \cos(\varphi+\pi)\end{bmatrix} = -\mathbf{r}_1,$$

$$\mathbf{F}_1 = F_{x'}\begin{bmatrix}\cos(\varphi)\\ \sin(\varphi)\end{bmatrix}, \mathbf{F}_2 = F_{x'}\begin{bmatrix}\cos(\varphi+\pi)\\ \sin(\varphi+\pi)\end{bmatrix} = -\mathbf{F}_1.$$

We then obtain (note that $P_0$ signifies the power incident on one of the gratings in the spinner only):

Orbital torque, x-polarization:

(22) $$\tau_{orb} = \mathbf{r}_1 \times \mathbf{F}_1 + \mathbf{r}_2 \times \mathbf{F}_2 = 2r\frac{n}{c_0}P_0\sum_i\left(p_i^2\cos^2(\varphi) + s_i^2\sin^2(\varphi)\right)\sin(\theta_i)|\cos(\theta_i)| =$$
$$2r\frac{n}{c_0}P_0\sum_i\left(T_{p,i}\cos^2(\varphi) + T_{s,i}\sin^2(\varphi)\right)\sin(\theta_i)$$

Orbital torque, LCP:

(23) $$\tau_{orb} = \mathbf{r}_1 \times \mathbf{F}_1 + \mathbf{r}_2 \times \mathbf{F}_2 = r\frac{n}{c_0}P_0\sum_i\left(T_{p,i} + T_{s,i}\right)\sin(\theta_i).$$

Spin torque, x-polarization:

(24) $\tau_{spin} = \tau_{spin}^1 + \tau_{spin}^2 = \frac{2}{n\omega} P_0 \sin(2\varphi) \sum_i p_i\, s_i \sin(\psi_i) \cos(\theta_i) |\cos(\theta_i)|.$

Spin torque, LCP

(25) $\tau_{spin} = \tau_{spin}^1 + \tau_{spin}^2 = -\frac{2}{n\omega} P_0 \{1 - \sum_i p_i\, s_i \cos(\psi_i) \cos(\theta_i) |\cos(\theta_i)|\}.$

The total torque is given by the sum of the orbital and spin contributions:

(26) $\tau_{\hat{z}} = \tau_{orb} + \tau_{spin}$

It is also useful to calculate the average torques over one full rotation cycle, $\varphi = 0 \rightarrow 2\pi$:

(27) $\frac{1}{2\pi} \int_0^{2\pi} \tau_{orb}^{x-pol}(\varphi)\, d\varphi = \frac{1}{2\pi} \int_0^{2\pi} \tau_{orb}^{LCP}(\varphi)\, d\varphi = r \frac{n}{c_0} P_0 \sum_i \left(T_{p,i} + T_{s,i}\right) \sin(\theta_i)$

(28) $\frac{1}{2\pi} \int_0^{2\pi} \tau_{spin}^{x-pol}(\varphi)\, d\varphi = 0$

(29) $\frac{1}{2\pi} \int_0^{2\pi} \tau_{spin}^{LCP}(\varphi)\, d\varphi = -\frac{2}{n\omega} P_0 \{1 - \sum_i p_i\, s_i \cos(\psi_i) \cos(\theta_i) |\cos(\theta_i)|\}$

**Forces and torques acting on a metaspinner in a unidirectional intensity gradient**

We consider the same RH metaspinner as before, but now it is situated in a field $I(x)$ that slowly varies in intensity along *x*. We approximate the intensity that hits the two metagratings with the intensity at their respective center positions $\mathbf{r}_1$ and $\mathbf{r}_2$. The positions along the x-axis are given by the instantaneous angles, $\varphi_1 = \varphi$ and $\varphi_2 = \varphi_1 + \pi$, and the moment arm $r$, which yields:

(30) $I(x_1) = I(x_{c.o.m}) - r \cdot \sin(\varphi) \frac{dI}{dx}|_{x_{c.o.m}}, I(x_2) = I(x_{c.o.m}) + r \cdot \sin(\varphi) \frac{dI}{dx}|_{x_{c.o.m}},$

where $x_{c.o.m}$ is the *x*-position of the metaspinner center of mass (c.o.m.). The net force acting on the c.o.m. is given by the vectorial sum of the two translation forces, i.e. $\mathbf{F}_{c.o.m.} = \mathbf{F}_1 + \mathbf{F}_2$. The forces are in turn proportional to the metagrating area $A$ and the local intensities. We then have for the LCP case:

(31) $\mathbf{F}_{c.o.m.} = \mathbf{F}_1 + \mathbf{F}_2 = -I(x_1) A C_{LCP} \begin{bmatrix} \cos(\varphi) \\ \sin(\varphi) \end{bmatrix} - I(x_2) A C_{LCP} \begin{bmatrix} \cos(\varphi + \pi) \\ \sin(\varphi + \pi) \end{bmatrix},$

where:

(32) $C_{LCP} = \frac{n}{2c_0} \sum_i \left(T_{p,i} + T_{s,i}\right) \sin(\theta_i).$

Inserting the intensities yields:

(33) $\mathbf{F}_{c.o.m.} = 2 A C_{LCP} r \sin(\varphi) \begin{bmatrix} \cos(\varphi) \\ \sin(\varphi) \end{bmatrix} \frac{dI}{dx}|_{x_{c.o.m}} = A C_{LCP} r \left\{ \begin{bmatrix} 0 \\ 1 \end{bmatrix} + \begin{bmatrix} \sin(2\varphi) \\ -\cos(2\varphi) \end{bmatrix} \right\} \frac{dI}{dx}|_{x_{c.o.m}}$

The intensity gradient in the x-direction thus gives rise to two distinct contributions to the force, where the first is independent of orientation $\varphi$ and always points perpendicular to the intensity gradient. We term this component the *transverse gradient force,* $\mathbf{F}_{trans}$. For the RH metaspinner and $dI/dx > 0$, $\mathbf{F}_{trans}$ points in the positive y-direction while a LH metaspinner would experience a force in the opposite direction. The direction of $\mathbf{F}_{trans}$ is thus determined by the handedness of the spinner and, thereby, by its spinning direction. As illustrated in Fig. S11, the result is that a metaspinner situated at some fixed radial distance $R_{c.o.m.} > r$ from the center of a Gaussian beam will orbit the beam center in the opposite direction to its spinning motion. $\mathbf{F}_{trans}$ will, in this special case, manifest as a fictious *gradient orbital torque* with magnitude $AC_{LCP} r R_{c.o.m.} \frac{dI}{dx}|_{R_{c.o.m.}}$.

The second force component in Eq. (33) also has constant magnitude, but it varies with orientation $\varphi$ such that a circular movement of the c.o.m. is induced. However, it does not result in a net displacement if averaged over all angles. The net result of the two force components is thus that the metaspinner will move in a spiral pattern in the direction perpendicular to the intensity gradient.

In addition to $\mathbf{F}_{c.o.m.}$ above, the metaspinner is subject to the classical gradient force, $\mathbf{F}_{grad}$, which always points in the direction of the intensity gradient, here along *x*. $\mathbf{F}_{grad}$ does not appear in the derivation above, however, and has to be added ad hoc.

We can use Eq. (33) also for x-polarized incidence, but we need to replace $C_{LCP}$ with

(34) $$C_{x-pol} = \frac{n}{c_0} \sum_i \left(T_{p,i} \cos^2(\varphi) + T_{s,i} \sin^2(\varphi)\right) \sin(\theta_i).$$

The magnitude of $\mathbf{F}_{trans}$ will now fluctuate with $\varphi$, but it will remain positive since $C_{x-pol}$ does not change sign as $\varphi$ varies. Similarly, the angular variation of $C_{x-pol}$ will in general cause the second force component in Eq. (33) to drive an elliptical rather than a circular motion.

The net torques relative to the metaspinner c.o.m., located at $\mathbf{r}_{c.o.m.}$, can be obtained in a similar fashion as above, but they do not depend on the intensity gradient. The results are thus simply:

Orbital torque, x-polarization:

(35) $$\tau_{orb} = (\mathbf{r}_1 - \mathbf{r}_{c.o.m.}) \times \mathbf{F}_1 + (\mathbf{r}_2 - \mathbf{r}_{c.o.m.}) \times \mathbf{F}_2 = 2rI(x_{c.o.m})AC_{x-pol}.$$

Orbital torque, LCP:

(36) $$\tau_{orb} = (\mathbf{r}_1 - \mathbf{r}_{c.o.m.}) \times \mathbf{F}_1 + (\mathbf{r}_2 - \mathbf{r}_{c.o.m.}) \times \mathbf{F}_2 = 2rI(x_{c.o.m})AC_{LCP}.$$

Spin torque, x-polarization:

(37) $$\tau_{spin} = \tau_{spin,1} + \tau_{spin,2} = \frac{2}{n\omega} I(x_{c.o.m}) A \sin(2\varphi) \sum_i p_i\, s_i \sin(\psi_i) \cos(\theta_i) |\cos(\theta_i)|.$$

Spin torque, LCP

(38) $$\tau_{spin} = \tau_{spin,1} + \tau_{spin,2} = -\frac{2}{n\omega} I(x_{c.o.m}) A\{1 - \sum_i p_i\, s_i \cos(\psi_i) \cos(\theta_i)\, |\cos(\theta_i)|\}.$$